\documentclass[12pt,draftclsnofoot,letterpaper,oneside,onecolumn]{IEEEtran}
\def\BibTeX{{\rm B\kern-.05em{\sc i\kern-.025em b}\kern-.08em
    T\kern-.1667em\lower.7ex\hbox{E}\kern-.125emX}}
\usepackage{amsmath}
\usepackage{amsthm}
\usepackage{amsfonts}
\usepackage{amssymb}
\usepackage{graphicx}
\usepackage{cite}
\usepackage{subfigure}
\usepackage{balance}
\usepackage{algorithm}
\usepackage{algorithmic}
\usepackage{enumerate}
\usepackage{color}
\usepackage{array}
\usepackage{indentfirst}
\usepackage{multirow}
\usepackage{booktabs}
\usepackage{color}
\usepackage{slashbox}
\usepackage{diagbox}
\usepackage{threeparttable}
\usepackage[svgnames,table]{xcolor}
\usepackage{bm}
\usepackage{bbm}
\usepackage{subfigure}
\usepackage{epstopdf}

\makeatletter

\newcommand{\Rmnum}[1]{\expandafter\@slowromancap\romannumeral #1@}
\makeatother

\begin{document}
\title{ Scalable Predictive Beamforming for IRS-Assisted Multi-User Communications: \\ A Deep Learning Approach \vspace{-0.2cm} }


\author{\IEEEauthorblockN{Chang Liu, \emph{Member, IEEE}, Xuemeng Liu, Zhiqiang Wei, \emph{Member, IEEE}, \\ Derrick Wing Kwan Ng, \emph{Fellow, IEEE}, and Robert Schober, \emph{Fellow, IEEE} \vspace{-0.2cm}}

\thanks{C. Liu and D. W. K. Ng are with the University of New South Wales, Sydney, Australia. X. Liu is with the University of Sydney, Sydney, Australia. Z. Wei is with Xi'an Jiaotong University, Xi'an, China. C. Liu and R. Schober are with Friedrich-Alexander-University Erlangen-N$\ddot{\mathrm{u}}$rnberg, Germany.
This work has been presented in part at the IEEE Global Communications Conference (GLOBECOM), 2021 \cite{liu2021deeplearningconference}.
}

}

\maketitle

\vspace{-1.65cm}

\begin{abstract} \vspace{-0.35cm}
Beamforming design for intelligent reflecting surface (IRS)-assisted multi-user communication (IRS-MUC) systems critically depends on the acquisition of accurate channel state information (CSI).
However, channel estimation (CE) in IRS-MUC systems causes a large signaling overhead for training due to the large number of IRS elements.
In this paper, taking into account user mobility, we adopt a deep learning (DL) approach to implicitly learn the historical line-of-sight (LoS) channel features and predict the IRS phase shifts to be adopted for the next time slot for maximization of the weighted sum-rate (WSR) of the IRS-MUC system.
With the proposed predictive approach, we can avoid full-scale CSI estimation and facilitate low-dimensional CE for transmit beamforming design such that the signaling overhead is reduced by a scale of $\frac{1}{N}$, where $N$ is the number of IRS elements.
To this end, we first develop a universal DL-based predictive beamforming (DLPB) framework featuring a two-stage predictive-instantaneous beamforming mechanism.
As a realization of the developed framework, a location-aware convolutional long short-term memory (CLSTM) graph neural network (GNN) is developed to facilitate effective predictive beamforming at the IRS, where a CLSTM module is first adopted to exploit the spatial and temporal features of the considered channels and a GNN is then applied to empower the designed neural network with high scalability and generalizability.
Furthermore, in the second stage, based on the predicted IRS phase shifts, an instantaneous CSI-aware fully-connected neural network is designed to optimize the transmit beamforming at the access point.
Simulation results demonstrate that the proposed framework not only achieves a better WSR performance and requires a lower CE overhead compared with state-of-the-art benchmarks, but also is highly scalable in the numbers of users.
\vspace{-0.55cm}
\end{abstract}


\begin{IEEEkeywords} \vspace{-0.45cm}
Intelligent reflecting surface, predictive beamforming, deep learning, graph neural network.
\end{IEEEkeywords}

\newpage
\section{Introduction\label{sect: intr}}
Recently, intelligent reflecting surfaces (IRSs), also known as reconfigurable intelligent surfaces \cite{yu2021smart, liu2021reconfigurable}, have been proposed as a promising smart radio technology for enhancing future communication systems.
Thanks to breakthroughs in the development of passive and programmable metamaterials, IRSs have the powerful capability of passively customizing the wireless channels between transmitters and receivers to improve the system data rate and communication reliability in a low-cost, light-weight, and low-energy-overhead manner \cite{gong2020toward, li2021many}.
Generally, an IRS consists of a large number of passive and reconfigurable reflecting elements \cite{wu2021intelligent}, where each element is reconfigurable and can be controlled independently.
In particular, by adapting the phase shifts of the reflecting elements according to the available channel state information (CSI), an IRS can alter the reflection directions of incident signals towards the desired receivers \cite{di2020smart}.
Therefore, it is expected that IRSs will be widely adopted and deployed to assist wireless systems in enhancing communication efficiency, and the study of IRS has attracted significant attention from both academia and industry \cite{pan2021reconfigurable, yu2021convolutional, Samsung6G}.

To fully unleash the potential of IRSs, various effective beamforming schemes have been proposed for IRS-assisted communications \cite{li2021intelligent, yuan2021reconfigurable}.
For example, a joint active and passive beamforming scheme \cite{wu2019beamforming}  was proposed  to minimize the total transmit power for IRS-assisted communication networks.
Subsequently, the authors of \cite{guo2020weighted} developed a low-complexity fractional programming (FP)-based method to maximize the weighted sum-rate (WSR) of an IRS-aided multiuser downlink multiple-input single-output (MISO) system.
Furthermore, the application of deep learning (DL)-based methods to IRS systems has been widely studied.
Indeed, by exploiting powerful data-driven methods, DL-based schemes generally achieve a better system performance compared with traditional model-based schemes \cite{yuan2021reconfigurable, wu2019beamforming, guo2020weighted, hu2021robust, cai2022resource}, showing the effectiveness of DL for IRS-assisted systems \cite{wang2021interplay}.
For instance, the authors of \cite{huang2020reconfigurable} studied a deep deterministic policy gradient (DDPG)-based approach for IRS-assisted systems and proposed a DDPG-empowered joint passive and active beamforming scheme to maximize the sum-rate of IRS-assisted multiuser downlink MISO systems. The proposed scheme can achieve a performance comparable to that of the state-of-the-art benchmarks requiring a relatively low computational complexity.
Moreover, an unsupervised DL-based phase-shift optimization method was proposed in \cite{gao2020unsupervised}. In particular, the developed scheme achieves a good rate performance while incurring only a relatively small computational overhead.
The aforementioned beamforming designs, i.e., \cite{wu2019beamforming, guo2020weighted, wang2021interplay, huang2020reconfigurable, gao2020unsupervised}, are based on the idealistic assumption that perfect channel estimation (CE) is available.
However, this assumption is generally not valid in practice since practical CE schemes \cite{zheng2022survey, liu2022deepresidual} for acquiring the CSI in IRS-assisted systems cause inevitable estimation errors and introduce a high signaling overhead.
For example, the authors of \cite{jensen2020optimal} developed a discrete Fourier transform-based CE scheme, which, however, requires a pilot sequence length of no less than the number of IRS elements. Such high signaling overhead cannot satisfy even the basic latency requirements for high mobility scenarios \cite{gong2020toward, wu2021intelligent}.
To reduce the signaling overhead, various effective approaches, such as IRS grouping \cite{zheng2020intelligent} and compressed sensing \cite{he2019cascaded}, were proposed. Although these methods can reduce the CE overhead to a certain extent, they reduce the CE accuracy, which may in turn lead to system performance degradation.
Therefore, existing CE schemes either require an exceedingly large signaling overhead or suffer from a limited CE accuracy.
Thus, efficient CSI acquisition remains a major obstacle for IRS deployment in practice \cite{pan2021reconfigurable, Samsung6G}.

Based on the above discussion, one important technical challenge for realizing IRS-assisted communication networks is how to design practical beamforming schemes considering the required CE overhead for time-varying channel environments \cite{wu2021intelligent, di2020smart, pan2021reconfigurable}.
If the IRS phase shifts could be somehow preset and fixed, the CSI acquisition problem would boil down to a low-dimensional MISO CE problem, i.e., only the CSI of the access point (AP)-IRS-user links would have to be acquired, thus significantly reducing the CE training overhead.
More importantly, existing CE approaches designed for MISO systems, e.g., \cite{zheng2022survey, liu2022deepresidual, tse2005fundamentals}, could be directly applied to such IRS-assisted networks simplifying the practical implementation of IRS.
Inspired by this observation, in this paper, we take into account user mobility and propose a learning-based predictive beamforming scheme to implicitly learn the historical line-of-sight (LoS) channel features for predicting the IRS phase shifts for maximization of the average WSR of an IRS-assisted multi-user communication (IRS-MUC) system.
In particular, we note that traditional neural networks \cite{liu2020deeptransfer, lxm2020deepresidual, yuan2020learning, liu2020location, xie2019activity, xie2020unsupervised, xie2020deep} need to be retrained once the number of users changes.
To further reduce such neural network retraining overhead, we exploit the emerging graph neural network (GNN) technology \cite{wu2020comprehensive} to design a scalable predictive beamforming scheme which takes into account the heterogeneous mobilities of the users.
Specifically, GNNs are designed to process structured data to learn the characteristics of possibly sophisticated data relationships and interdependencies, e.g., the interference between users, for improving the system performance.
In fact, some preliminary works, e.g., \cite{shen2020graph, eisen2020optimal, lee2020graph}, have confirmed the excellent scalability and generalization capability of GNNs for wireless resource allocation and management.
By exploiting GNNs' property of permutation invariance/equivariance \cite{shen2020graph}, a well-trained GNN can be directly adopted for beamforming design in IRS-MUC systems for arbitrary numbers of users without the need for retraining, thus realizing a scalable predictive beamforming scheme.
The main contributions of this work are as follows:
\begin{enumerate}[(1)]
\item We design a predictive communication protocol and formulate a general predictive beamforming problem for maximization of the average WSR of IRS-MUC systems. By presetting the IRS with the predicted phase shifts, we can replace the estimation of the large-dimensional full instantaneous CSI (ICSI) with a conventional small-dimensional estimation of the end-to-end AP-IRS-user MISO channel.
    This approach not only yields an excellent communication rate performance, but also reduces the CE signaling overhead by a factor of $\frac{1}{N}$ ($N$ is the number of IRS elements) compared to full ICSI estimation.

\item To solve the formulated problem, a DL-based predictive beamforming (DLPB) framework is developed, which decomposes the sophisticatedly coupled original design problem into two sub-problems: (a) the LoS-aware predictive IRS beamforming design and (b) the ICSI-aware instantaneous transmit beamforming design. Accordingly, a two-stage deep neural network (DNN) training mechanism is designed to facilitate (a) and (b) and to enhance the practicality of the proposed approach.

\item As a realization of the developed framework, a location-aware convolutional long short-term memory (CLSTM) GNN (LA-CLGNN) is designed for predictive IRS beamforming, where a CLSTM module is adopted to extract the spatial-temporal features of the historical LoS channels to improve the learning performance and a GNN is exploited to enhance the scalability of the LA-CLGNN. Specifically, each node in the GNN is trained with the same parameters such that the well-trained GNN is applicable for systems with an arbitrary number of users without the need for retraining. Furthermore, an ICSI-aware fully-connected neural network (IA-FNN) is proposed for the beamforming design at the AP.


\item Extensive simulation results are provided to demonstrate the excellent performance of the proposed framework in terms of WSR performance, scalability, and generalizability, respectively. In particular, the proposed predictive scheme is shown to be scalable and to achieve a significant performance gain compared with the traditional FP-based method \cite{guo2020weighted} requiring perfect full ICSI.
\end{enumerate}

The remainder of this paper is organized as follows.
In Section II, we introduce the IRS-MUC system model and propose a predictive transmission protocol for the considered system model.
In Section III, we formulate predictive beamforming as an optimization problem.
To solve the formulated problem, a DL-based predictive beamforming framework for IRS-MUC systems is proposed in Section IV.
As a realization of the developed framework, in Section V, an LA-CLGNN-based IRS predictive beamforming design and an IA-FNN-based instantaneous AP beamforming design are developed, and a DL-based predictive beamforming algorithm is proposed.
Subsequently, Section VI reports extensive simulation results to verify the
effectiveness of the proposed framework, and Section VII concludes this paper.

\emph{Notations}:
Superscripts $H$ and $T$ are used to represent conjugate transpose and transpose, respectively.
$\mathbb{N}_1$, $\mathbb{R}$, and $\mathbb{C}$ represent the sets of natural, real, and complex numbers, respectively.
$\arccos(\cdot)$ denotes the inverse cosine function and $\mathcal{U}(a,b)$ is the uniform distribution within $[a,b]$. $\|\cdot\|_F$, $\|\cdot\|$, and $|\cdot|$ refer to the Frobenius norm, $\ell_2$ norm, and absolute value, respectively.
${\mathcal{CN}}( \bm{\mu},\mathbf{\Sigma} )$ and ${\mathcal{N}}( \bm{\mu},\mathbf{\Sigma} )$ are used to denote the circularly symmetric complex Gaussian (CSCG) and real-valued Gaussian distributions, where $\bm{\mu}$ and $\mathbf{\Sigma}$ denote the mean vector and the covariance matrix, respectively.
$\otimes$ indicates the Kronecker product.
$\mathbb{E}\{\cdot\}$ denotes the statistical expectation operation and $\mathrm{diag}(\cdot)$ represents a diagonal matrix with the input vector as the diagonal elements.
In addition, we use $\mathrm{Re}\{\cdot\}$ and $\mathrm{Im}\{\cdot\}$ to denote the real and imaginary parts of a complex-valued matrix, respectively.
$\mathbf{x}(a:b)$ denotes the elements of a vector $\mathbf{x}$ ranging from the $a$-th element to the $b$-th element of $\mathbf{x}$.
$[\cdot]_i$ represents the $i$-th element of a vector.
$\mathbf{X}(:,a)$ denotes the $a$-th column of matrix $\mathbf{X}$.

\begin{figure}[t]
  \centering
  \includegraphics[width=0.46\linewidth]{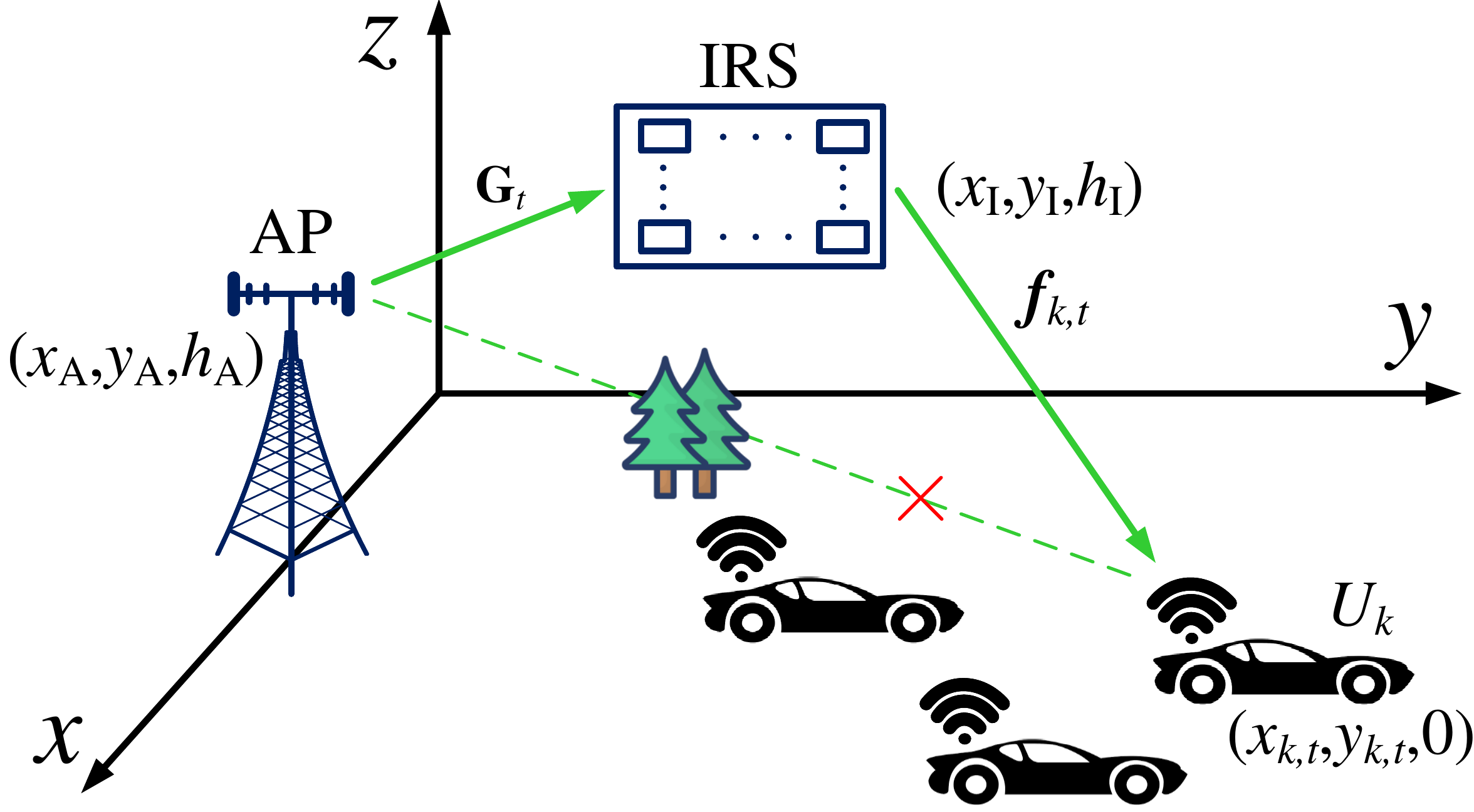}\vspace{-0.8cm}
  \caption{ The downlink of the considered IRS-MUC system. }\vspace{-1.0cm}\label{Fig:scenario}
\end{figure}

\vspace{-0.4cm}
\section{System Model}
In this paper, a typical downlink IRS-MUC system is considered which consists of one AP, one IRS, and $K$ users. As shown in Fig. \ref{Fig:scenario}, the AP is equipped with an $M$-element uniform linear array to serve the $K$ single-antenna users with the assistance of an IRS planar array consisting of $N = N_y \times N_z$ passive reflecting elements, where $N_y$ and $N_z$ denote the number of elements in the $y$- and $z$-directions, respectively.
In the sequel, we adopt subscripts $k$ and $t$ as indices for user $k$, $k \in \mathcal{K} \triangleq \{1,2,\cdots,K\}$,  denoted by $U_k$, and time slot $t$, $t \in \mathbb{N}_1$, respectively. \vspace{-0.4cm}

\vspace{-0.2cm}
\subsection{Mobility Model}
\begin{figure}[t]
  \centering
  \includegraphics[width=0.65\linewidth]{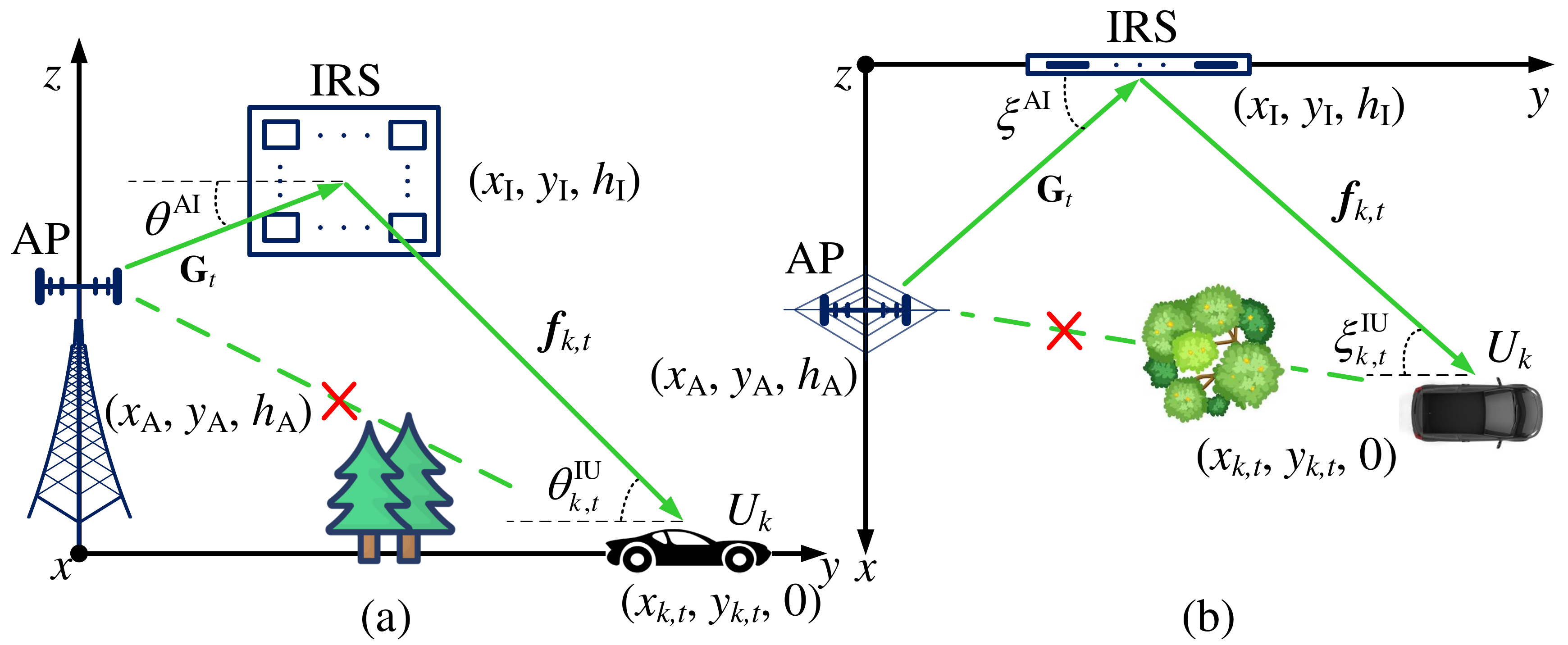}\vspace{-0.8cm}
  \caption{ The downlink of the considered IRS-MUC system with $K=1$ for illustration.
  (a) Side view; (b) Bird's-eye view.}\label{Fig:scenario_vertical_horizontal} \vspace{-1.0cm}
\end{figure}

In this work, we take into account user mobility and focus on a dynamic scenario where each user is in motion while the AP and the IRS are static.
As illustrated in Fig. \ref{Fig:scenario_vertical_horizontal}, we adopt a three-dimensional (3D) Cartesian coordinate system to define locations. The locations of AP and IRS are given by $\mathbf{L}_\mathrm{A} = [x_\mathrm{A},y_\mathrm{A},h_\mathrm{A}]^T$ and $\mathbf{L}_\mathrm{I} = [x_\mathrm{I},y_\mathrm{I},h_\mathrm{I}]^T$, respectively, where $x_\mathrm{A}$, $y_\mathrm{A}$, and $h_\mathrm{A}$ ($x_\mathrm{I}$, $y_\mathrm{I}$, and $h_\mathrm{I}$) are the $x$-, $y$-, and $z$-coordinates of AP (IRS), respectively.
Moreover, the location of $U_k$ in time slot $t$ is $\mathbf{L}_{k,t} = [x_{k,t},y_{k,t},0]^T$, where $x_{k,t}$ and $y_{k,t}$ are the $x$- and $y$-coordinates, respectively.
Accordingly, the AP-to-IRS and IRS-to-$U_k$ distances in time slot $t$ can be expressed as
$d^\mathrm{AI} = \sqrt{(x_\mathrm{A} - x_\mathrm{I})^2 + (y_\mathrm{A} - y_\mathrm{I})^2 + (h_\mathrm{A} - h_\mathrm{I})^2}$ and $d^\mathrm{IU}_{k,t} = \sqrt{ (x_\mathrm{I} - x_{k,t})^2 + (y_\mathrm{I} - y_{k,t})^2 + h_\mathrm{I}^2 }$, respectively.
Without loss of generality, the kinematic equation of $U_k$ can be formulated as \cite{liu2022learning, li2022novel}
\begin{equation}\label{movement_function}
  \mathbf{L}_{k,t+1} = \mathbf{L}_{k,t} + \bar{\mathbf{v}}_{k,t} \Delta T + \mathbf{u}_{k,t}, \forall k,t.
\end{equation}
Here, $\bar{\mathbf{v}}_{k,t} = [\bar{v}_{k,t}^x, \bar{v}_{k,t}^y, 0]^T$ denotes the average velocity vector of $U_k$ in time slot $t$, where $\bar{v}_{k,t}^x$ and $\bar{v}_{k,t}^y$ represent the velocity projections on the $x$- and $y$-axes, respectively.
Both the magnitude $a_{k,t} = \|\bar{\mathbf{v}}_{k,t}\|$ and the direction $b_{k,t} = \mathrm{arccos} (\frac{\bar{v}_{k,t}^x}{\|\bar{\mathbf{v}}_{k,t}\|})$ are assumed to follow uniform distributions, i.e., $a_{k,t} \sim \mathcal{U}(A_1,A_2)$ and $b_{k,t} \sim \mathcal{U}(B_1,B_2)$, where $A_1$, $A_2$, $B_1$, $B_2$ are constants with $A_1 \leq A_2$ and $B_1 \leq B_2$ \cite{9076668, liu2014maximum}.
In addition, $\Delta T$ denotes the duration of a time slot\footnotemark\footnotetext{$\Delta T$ is chosen sufficiently short such that $\mathbf{L}_{k,t}$ can be assumed to be constant during $\Delta T$ \cite{liu2022learning, 9076668}. }, and $\mathbf{u}_{k, t} = [u_{k,t}^x, u_{k,t}^y, u_{k,t}^z]^T$ is adopted to characterize environmental uncertainties caused by the measurement, where $u_{k,t}^\varepsilon \sim \mathcal{N}(0,\sigma_u^2)$, $\varepsilon = \{x,y,z\}$, denotes the uncertainty offset in $\varepsilon$-direction and $\sigma_u^2$ is the uncertainty variance.
Specifically, for every time instant, we assume that only the historical locations are known at the AP and this information will be exploited by the proposed DL framework for predictive IRS beamforming design.

\vspace{-0.4cm}
\subsection{Channel Model}

Due to unfavorable signal blockages, the direct AP-to-user links are assumed to be shadowed and are neglected in the following, as illustrated in Fig. \ref{Fig:scenario}, as is commonly adopted in e.g., \cite{huang2020reconfigurable, you2020energy}.
Without loss of generality, a Rician fading model taking into account both the LoS component and the non-LoS (NLoS) component is adopted for all communication channels \cite{wu2019beamforming, guo2020weighted}.
Thus, the channel of the AP-to-IRS link in time slot $t$ is modelled as
\begin{equation}\label{G_model}
  \mathbf{G}_t = \sqrt{\frac{\alpha^{\mathrm{AI}}\beta^{\mathrm{AI}}}{\beta^{\mathrm{AI}} + 1}}\bar{\mathbf{G}} + \sqrt{\frac{\alpha^{\mathrm{AI}}}{\beta^{\mathrm{AI}} + 1}}\tilde{\mathbf{G}}_t.
  \vspace{-0.05cm}
\end{equation}
Here, $\alpha^{\mathrm{AI}}$ and $\beta^\mathrm{AI}$ denote the path loss and the Rician factor of the AP-to-IRS link, respectively. $\tilde{\mathbf{G}}_t \in \mathbb{C}^{N \times M}$ is the NLoS component of $\mathbf{G}_t$.
To illustrate the relative locations of AP, IRS, and users, Fig. \ref{Fig:scenario_vertical_horizontal} depicts the side view and the bird's-eye view for the downlink of the considered IRS-MUC system.
According to \cite{tse2005fundamentals, liu2016blind, liu2015blind}, the LoS component of $\mathbf{G}_t \in \mathbb{C}^{N\times M}$ can be written as \vspace{-0.2cm}
\begin{equation}\label{G_bar}
  \bar{\mathbf{G}} = \mathbf{a}^\mathrm{IRS}_y(\theta^\mathrm{AI}, \xi^\mathrm{AI}) \otimes \mathbf{a}^\mathrm{IRS}_z(\theta^\mathrm{AI}, \xi^\mathrm{AI}) \otimes (\mathbf{a}^\mathrm{AP}(\theta^\mathrm{AI}, \xi^\mathrm{AI}))^H,
\end{equation}
where \vspace{-0.2cm}
\begin{equation}\label{}
  \mathbf{a}^\mathrm{AP}(\theta^\mathrm{AI}, \xi^\mathrm{AI}) = [   1, e^{-j\frac{2\pi\Delta d_\mathrm{A}}{\lambda_c} \sin(\theta^\mathrm{AI})\cos(\xi^\mathrm{AI})}, \cdots,
   e^{-j\frac{2\pi (M -1) \Delta d_\mathrm{A}}{\lambda_c} \sin(\theta^\mathrm{AI})\cos(\xi^\mathrm{AI})}]^T ,
\end{equation}
\begin{equation}\label{}
  \mathbf{a}^\mathrm{IRS}_y(\theta^\mathrm{AI}, \xi^\mathrm{AI}) = [   1, e^{-j\frac{2\pi\Delta d_{\mathrm{I}y}}{\lambda_c} \sin(\theta^\mathrm{AI})\cos(\xi^\mathrm{AI})}, \cdots,   \\
   e^{-j\frac{2\pi (N_y -1) \Delta d_{\mathrm{I}y}}{\lambda_c} \sin(\theta^\mathrm{AI})\cos(\xi^\mathrm{AI})}]^T ,
\end{equation}
and \vspace{-0.2cm}
\begin{equation}\label{}
  \mathbf{a}^\mathrm{IRS}_z(\theta^\mathrm{AI}, \xi^\mathrm{AI}) = [    1, e^{-j\frac{2\pi\Delta d_{\mathrm{I}z}}{\lambda_c} \sin(\theta^\mathrm{AI})\sin(\xi^\mathrm{AI})}, \cdots,   \\
   e^{-j\frac{2\pi (N_z -1) \Delta d_{\mathrm{I}z}}{\lambda_c} \sin(\theta^\mathrm{AI})\sin(\xi^\mathrm{AI})}]^T
 \vspace{-0.2cm}
\end{equation}
are the steering vectors of the AP, IRS on the $y$-axis, and IRS on the $z$-axis, respectively. Here, parameters $\theta^\mathrm{AI}$ and $\xi^\mathrm{AI}$ represent the horizontal and vertical angles-of-arrival (AoAs) from the AP at the IRS, respectively.
According to Fig. \ref{Fig:scenario_vertical_horizontal}, we have $\sin(\theta^\mathrm{AI}) = \frac{|h_\mathrm{I} - h_\mathrm{A}|}{d^\mathrm{AI}}$, $\cos(\xi^\mathrm{AI}) = \frac{|y_\mathrm{I}|}{d^\mathrm{AI}}$, and $\sin(\xi^\mathrm{AI}) = \frac{|x_\mathrm{A}|}{d^\mathrm{AI}}$.
Besides, $\lambda_c$ denotes the operating wavelength of the considered system, $\Delta d_\mathrm{A}$ is the distance between two adjacent antenna elements of the AP, and $\Delta d_{\mathrm{I}y}$ and $\Delta d_{\mathrm{I}z}$ are the distances between two adjacent IRS elements in $y$- and $z$-directions, respectively.
Similarly, the channel of the IRS-to-$U_k$ link in time slot $t$ is given by \vspace{-0.2cm}
\begin{equation}\label{f_model}
  \mathbf{f}_{k,t} = \sqrt{\frac{\alpha^{\mathrm{IU}}_{k,t}\beta^{\mathrm{IU}}_k}{\beta^\mathrm{IU}_k + 1}}\bar{\mathbf{f}}_{k,t} + \sqrt{\frac{\alpha^{\mathrm{IU}}_{k,t}}{\beta^\mathrm{IU}_k + 1}}\tilde{\mathbf{f}}_{k,t}.
\end{equation}
Here, $\alpha^{\mathrm{IU}}_{k,t}$ and $\beta^\mathrm{IU}_k$ represent the path loss and the Rician factor of the IRS-to-$U_k$ link, respectively. $\tilde{\mathbf{f}}_{k,t} \in \mathbb{C}^{N \times 1}$ is the NLoS component of $\mathbf{f}_{k,t}$.
In addition, \vspace{-0.2cm}
\begin{equation}\label{f_bar}
  \bar{\mathbf{f}}_{k,t} = \mathbf{a}^\mathrm{IRS}_y(\theta^\mathrm{IU}, \xi^\mathrm{IU}) \otimes \mathbf{a}^\mathrm{IRS}_z(\theta^\mathrm{IU}, \xi^\mathrm{IU}) \vspace{-0.2cm}
\end{equation}
denotes the LoS component of $\mathbf{f}_{k,t}$, where \vspace{-0.2cm}
\begin{equation}\label{}
  \mathbf{a}^\mathrm{IRS}_y(\theta^\mathrm{IU}_{k,t}, \xi^\mathrm{IU}_{k,t}) = [   1, e^{-j\frac{2\pi\Delta d_{\mathrm{I}y}}{\lambda_c} \sin(\theta^\mathrm{IU}_{k,t})\cos(\xi^\mathrm{IU}_{k,t})}, \cdots,   \\
   e^{-j\frac{2\pi (N_y -1) \Delta d_{\mathrm{I}y}}{\lambda_c} \sin(\theta^\mathrm{IU}_{k,t})\cos(\xi^\mathrm{IU}_{k,t})}]^T
\end{equation}
and \vspace{-0.2cm}
\begin{equation}\label{}
  \mathbf{a}^\mathrm{IRS}_z(\theta^\mathrm{IU}_{k,t}, \xi^\mathrm{IU}_{k,t}) = [   1, e^{-j\frac{2\pi\Delta d_{\mathrm{I}z}}{\lambda_c} \sin(\theta^\mathrm{IU}_{k,t})\sin(\xi^\mathrm{IU}_{k,t})}, \cdots,   \\
   e^{-j\frac{2\pi (N_z -1) \Delta d_{\mathrm{I}z}}{\lambda_c} \sin(\theta^\mathrm{IU}_{k,t})\sin(\xi^\mathrm{IU}_{k,t})}]^T
\end{equation}
represent the steering vectors of the IRS with
$\sin(\theta^\mathrm{IU}_{k,t}) = \frac{|h_\mathrm{I}|}{d^\mathrm{IU}_{k,t}}$, $\cos(\xi^\mathrm{IU}_{k,t}) = \frac{|y_{k,t} - y_\mathrm{I}|}{d^\mathrm{IU}_{k,t}}$, and $\sin(\xi^\mathrm{IU}_{k,t}) = \frac{|x_{k,t}|}{d^\mathrm{IU}_{k,t}}$, as shown in Fig. \ref{Fig:scenario_vertical_horizontal}.

\vspace{-0.5cm}
\subsection{Signal Model}
The IRS can alter the reflection of incident signals by controlling its phase-shift matrix, $\mathbf{\Phi}_t = \mathrm{diag}(\mathbf{c}_t)\in \mathbb{C}^{N \times N}$, where $\mathbf{c}_t = [ e^{j\varphi_{1,t}}, e^{j\varphi_{2,t}},\cdots, e^{j\varphi_{N,t}} ]^T$ and $\varphi_{n,t}$ denotes the phase shift of the $n$-th IRS element, $n \in \mathcal{N} = \{1,2,\cdots,N\}$, in the time slot $t$.
Thus, the signal received at $U_k$ in the time slot $t$ is given by
\begin{equation}\label{r_kt}
  r_{k,t} = \mathbf{f}_{k,t}^H \mathbf{\Phi}_t \mathbf{G}_t\left(\sum_{i=1}^{K}\mathbf{w}_{i,t}s_{i,t}\right) + n_{k,t}, \forall t,k.
\end{equation}
Here, $\mathbf{w}_{i,t} \in \mathbb{C}^{M \times 1}$ and $s_{i,t} \sim \mathcal{CN}(0,1)$ are the beamforming vector and the transmitted data for user $U_i$ in the time slot $t$, respectively.
$n_{k,t} \sim \mathcal{CN}(0,\sigma_k^2)$ denotes the additive white Gaussian noise sample at $U_k$ in the time slot $t$ with $\sigma_k^2$ being the noise variance at the receiver of $U_k$.
Hence, the received signal-to-interference-plus-noise ratio (SINR) is obtained as
\begin{align}\label{SINR}
  \gamma_{k,t}(\mathbf{\Phi}_t,\mathbf{w}_{k,t}) = \frac{|\mathbf{f}_{k,t}^H\mathbf{\Phi}_t\mathbf{G}_t\mathbf{w}_{k,t}|^2}{\sum_{j \neq k}^K|\mathbf{f}_{k,t}^H\mathbf{\Phi}_t\mathbf{G}_t\mathbf{w}_{j,t}|^2 + \sigma_k^2}, \forall k,t.
\end{align}

\subsection{Transmission Protocol}

\begin{figure}[t]
  \centering
  \includegraphics[width=0.78\linewidth]{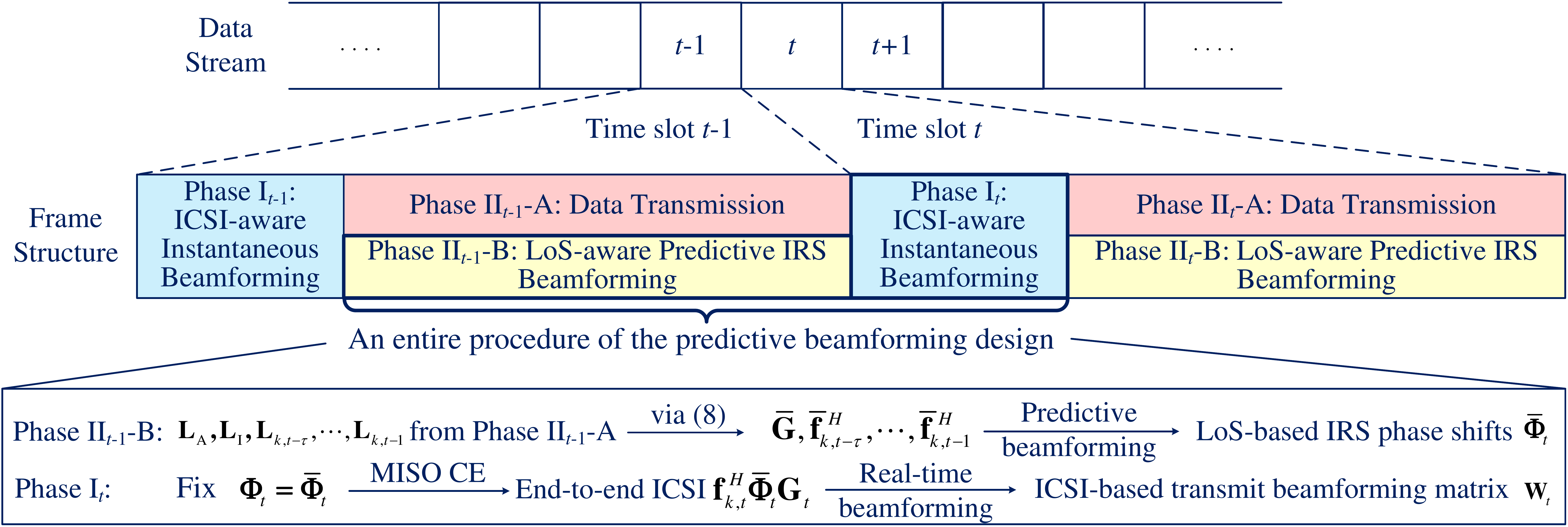}\vspace{-0.6cm}
  \caption{ The frame structure of the developed transmission protocol at the AP. }\label{Fig:tranmission protocol}
  \vspace{-1.0cm}
\end{figure}

Since reflecting elements can be fabricated at low cost, the number of IRS elements $N$ is expected to be very large, i.e., $N \geq 100$ \cite{yu2021smart, liu2021reconfigurable, gong2020toward, li2021many, wu2021intelligent, di2020smart}, to enable a significant passive beamforming gain.
Thus, the end-to-end cascaded IRS channel\footnotemark\footnotetext{Note that $\mathbf{f}_{k,t}^H\mathbf{\Phi}_t\mathbf{G}_t = \mathbf{c}_t^T\mathrm{diag}(\mathbf{f}_{k,t}^H) \mathbf{G}_t$, $\forall k,t$, holds.}, i.e., $\mathrm{diag}(\mathbf{f}_{k,t}^H) \mathbf{G}_t \in \mathbb{C}^{N \times M}$ is high dimensional, which incurs a large signaling overhead for CE in practice. However, since typically $N \gg M$, if the IRS phase shifts were known, the effective end-to-end ICSI would reduce to a low-dimensional MISO channel, i.e., $\mathbf{f}_{k,t}^H\mathbf{\Phi}_t\mathbf{G}_t \in \mathbb{C}^{1 \times M}$.
In this case, the signaling overhead required for estimating the end-to-end ICSI would be $1/N$ times lower than that required for estimating the full ICSI, i.e., $\mathrm{diag}(\mathbf{f}_{k,t}^H) \mathbf{G}_t \in \mathbb{C}^{N \times M}$, e.g., \cite{zheng2022survey, liu2022deepresidual, jensen2020optimal, zheng2020intelligent, he2019cascaded}.
Inspired by this observation, we propose a predictive transmission protocol with a hierarchical structure, where the IRS phase-shift matrix to be adopted in the next time slot is predicted and determined in advance such that only a MISO CE is required for designing the transmit beamforming at the AP.

The proposed transmission protocol is shown in Fig. \ref{Fig:tranmission protocol}. Each time slot $t$ consists of three phases: Phase I$_t$: ICSI-aware instantaneous beamforming, Phase II$_t$-A: Data transmission, and Phase II$_t$-B: LoS-aware predictive IRS beamforming.
Meanwhile, we assume that the channel coefficients and the user locations vary across time slots due to user mobility \cite{liu2022learning, tse2005fundamentals}, while they remain static within each time slot.
In time slot $t-1$, the predictive beamforming design for time slot $t$ involves (i) Phase II$_{t-1}$-B and (ii) Phase I$_{t}$.
In (i), the AP first acquires $\mathbf{L}_\mathrm{A}$, $\mathbf{L}_\mathrm{I}$, and the historical locations $\mathbf{L}_{k,t-\tau}, \cdots, \mathbf{L}_{k,t-1}$, $\forall k$, from the previous uplink transmissions and then assembles $\bar{\mathbf{G}}$ and the historical LoS channels $ \bar{\mathbf{f}}_{k,t-\tau}, \cdots,\bar{\mathbf{f}}_{k,t-1}$, $\forall k$, based on (\ref{G_bar}) and (\ref{f_bar}), respectively, where $\tau$ denotes the number of available historical time slots. Based on the historical LoS channels, the AP can predict and optimize the LoS-based IRS phase-shift matrix $\bar{\mathbf{\Phi}}_t$, which is employed in time slot $t$.
In (ii), the IRS first sets the phase-shift matrix as $\mathbf{\Phi}_t = \bar{\mathbf{\Phi}}_t$ for CE to acquire the end-to-end ICSI $\mathbf{f}_{k,t}^H\bar{\mathbf{\Phi}}_t\mathbf{G}_t$. Then, the estimated ICSI is adopted for optimization of transmit beamforming matrix $\mathbf{W}_t = [\mathbf{w}_{1,t},\mathbf{w}_{2,t},\cdots,\mathbf{w}_{K,t} ] $.
As will be shown in Section VI, the proposed protocol and design technology not only reduce the required CE overhead but also achieve a high beamforming performance.

\begin{figure}[t]
  \centering
  \includegraphics[width=0.7\linewidth]{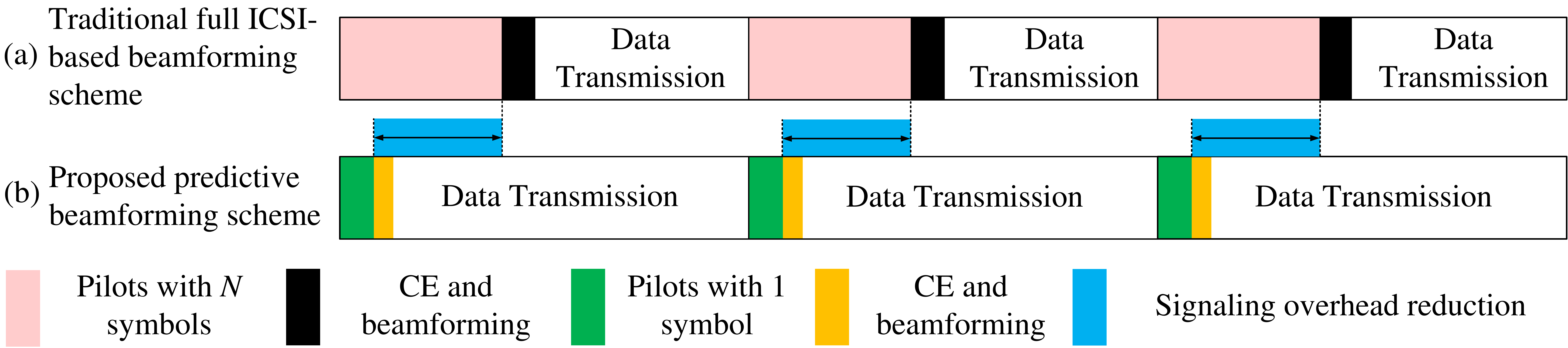}\vspace{-0.6cm}
  \caption{ Comparison of protocol structure for IRS-MUC systems with $K=1$ for illustration. }\label{Fig:CE_comparison}\vspace{-1.0cm}
\end{figure}

Furthermore, to provide more insight regarding the signaling overhead reduction for CE training, in Fig. \ref{Fig:CE_comparison}, we compare two schemes: (a) traditional full ICSI-based beamforming \cite{jensen2020optimal} and (b) the proposed predictive beamforming.
We observe that (a) incurs a large pilot signaling overhead, i.e., $N$ symbols, thus the remaining duration for data transmission may be limited.
In contrast, (b) exploits the proposed predictive scheme to bypass the need for full CSI estimation and reduces the pilot signaling overhead to 1 symbol as only MISO channels here to be estimated.
Thus, it is expected that the proposed predictive beamforming approach facilitates a much longer duration for data transmission compared with the conventional approaches.
This will be further verified in Section VI via simulations.
Furthermore, with the proposed approach, conventional CE schemes designed for MISO channels, e.g., \cite{zheng2022survey, liu2022deepresidual, tse2005fundamentals}, can be directly applied in IRS-MUC systems.

\vspace{-0.2cm}
\section{Problem Formulation}

In this paper, we aim to maximize the average achievable WSR via jointly optimizing the LoS-aware phase-shift matrix at the IRS and the ICSI-aware beamforming matrix at the AP subject to an instantaneous power constraint at the AP and the phase shift constraint at the IRS.
To this end, the optimization problem for time slot $t \in \mathcal{T} = \{t| t \geq \tau + 1, t \in \mathbb{N}_1 \}$ is formulated as follows
\begin{align}
\max_{\bar{\mathbf{\Phi}}_t} ~ &\mathbb{E}_{\bar{\mathbf{H}}_{t}|{\mathcal{H}}_t^{\tau}}\left\{\max_{{\mathbf{W}}_t}
\mathbb{E}_{\mathbf{H}_t|\hat{\mathbf{H}}_t} \left[\sum_{k = 1}^K\alpha_k\mathrm{log}_2\left(1 + \gamma_{k,t}(\bar{\mathbf{\Phi}}_t,\mathbf{w}_{k,t}) \right)\right]\right\} \label{P1} \\
\mathrm{s.t.}\,\,&\sum_{k = 1}^K\|\mathbf{w}_{k,t}\|^2\leq P, ~ 0 \leq \bar{\varphi}_{n,t} \leq 2\pi, \forall n \in \mathcal{N}, t \in \mathcal{T}. \notag
\end{align}
Here, $\alpha_k \ge 0$ is a weight for $U_k$, which can be used to prioritize users to enable different levels of quality-of-services (QoSs) \cite{guo2020weighted, wang2021interplay}, $P$ denotes the instantaneous maximum power budget at the AP and $\bar{\varphi}_{n,t}$ is the LoS-based phase shift of the $n$-th IRS element in the time slot $t$.
Besides, $\bar{\mathbf{\Phi}}_{t} = \mathrm{diag}([ e^{j\bar{\varphi}_{1,t}}, e^{j\bar{\varphi}_{2,t}},\cdots, e^{j\bar{\varphi}_{N,t}} ])$ and $\mathbf{W}_t$ as defined in Fig. \ref{Fig:tranmission protocol}, represent the LoS-based predictive phase-shift matrix and the ICSI-based beamforming matrix in time slot $t$, respectively.
Moreover, $\gamma_{k,t}(\cdot,\cdot)$ denotes the received SINR at $U_k$, as defined in (\ref{SINR}).
We note that (\ref{P1}) is a nested optimization problem. On the one hand, under the assumption of ICSI knowledge, the inner maximization is with respect to ${\mathbf{W}}_t$ for a given $\bar{\mathbf{\Phi}}_t$ obtained from the outer maximization.
In fact, the equivalent MISO channel of AP-IRS-$U_k$ link, denoted by $\mathbf{h}_{k,t}^H \triangleq \mathbf{f}_{k,t}^H\bar{\mathbf{\Phi}}_t\mathbf{G}_t \in \mathbb{C}^{1 \times M}$, can be estimated via existing CE methods, e.g., \cite{zheng2022survey, liu2022deepresidual, jensen2020optimal}, in Phase I of each time slot.
As such, the expectation $\mathbb{E}_{\mathbf{H}_t|\hat{\mathbf{H}}_t} $ is taken over all random realizations of  $\mathbf{H}_t \triangleq [\mathbf{h}_{1,t},\cdots,\mathbf{h}_{K,t}]$ given the estimated channels $\hat{\mathbf{H}}_t \triangleq [\hat{\mathbf{h}}_{1,t},\cdots,\hat
{\mathbf{h}}_{K,t}]$, where $\hat{\mathbf{h}}_{k,t} = \mathbf{h}_{k,t} + \Delta \mathbf{h}$ with $\Delta \mathbf{h} \in \mathbb{C}^{M \times 1}$ being a Gaussian random vector caused by the CE error.
On the other hand, assembling LoS channel knowledge, the outer maximization in (\ref{P1}) is with respect to predictive phase shift matrix $\bar{\mathbf{\Phi}}_{t}$ and the expectation $\mathbb{E}_{\bar{\mathbf{H}}_{t}|\mathcal{H}_t^{\tau}}\{\cdot\}$ is taken over all random realizations of $\bar{\mathbf{H}}_{t}$ given the historical LoS channels $\mathcal{H}_t^{\tau} \triangleq \{\bar{\mathbf{H}}_{t-1},\cdots,\bar{\mathbf{H}}_{t-\tau}\}$, where $\bar{\mathbf{H}}_{\lambda} \triangleq \{\bar{\mathbf{G}},\bar{\mathbf{f}}_{1,\lambda}, \cdots,\bar{\mathbf{f}}_{K,\lambda} \}$, $\forall \lambda \in \{t, t-1, \cdots, t-\tau\}$.
Generally, the problem in (\ref{P1}) is intractable due to the following reasons: i) a closed-form expression of the objective function is not available; ii) $\bar{\mathbf{\Phi}}_{t}$ and $\mathbf{W}_t$ are sophisticatedly coupled in the objective function.
To overcome these challenges, in the following section, we propose a learning-based beamforming framework to address problem (\ref{P1}).

\begin{figure}[t]
  \centering
  \includegraphics[width=0.8\linewidth]{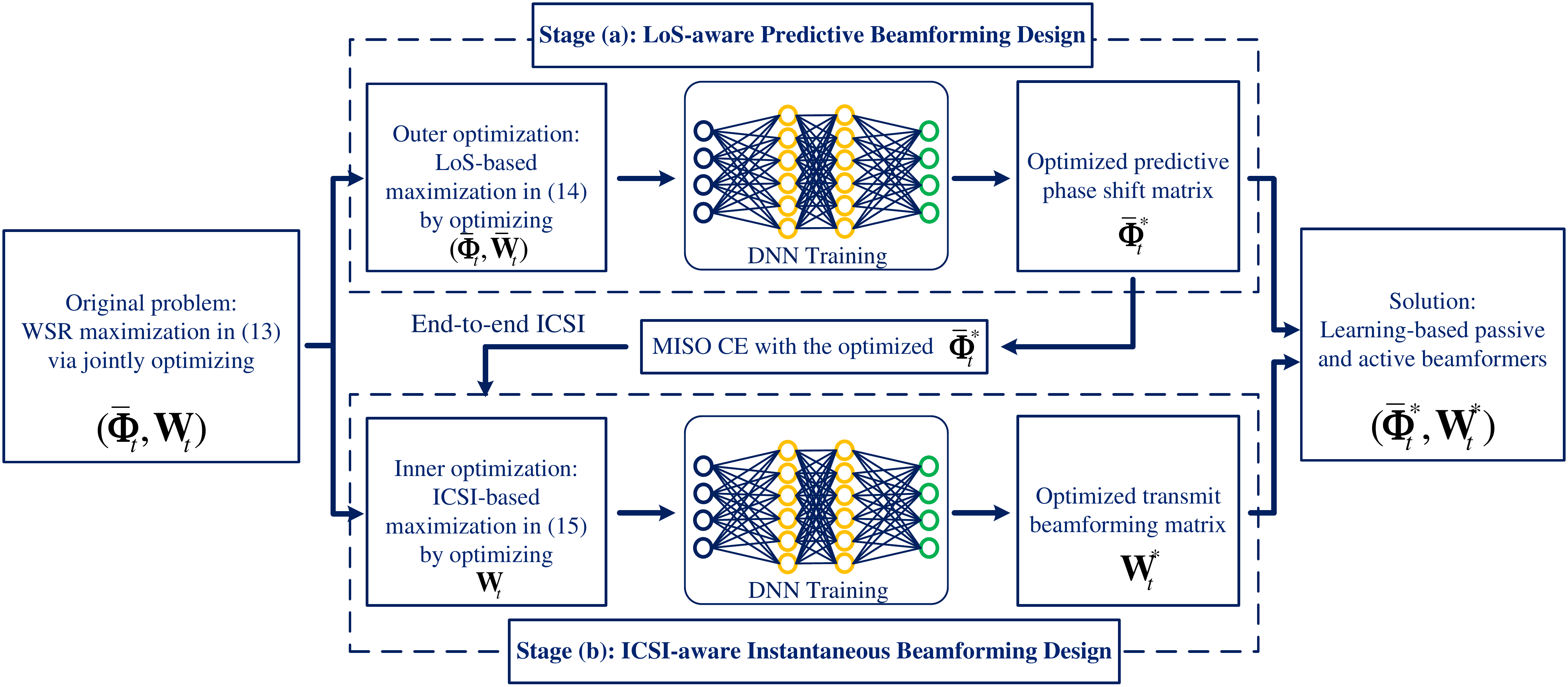}
  \vspace{-0.6cm}
  \caption{ The developed learning-based predictive beamforming framework for IRS-MUC systems. }\label{Fig:problem_solution}\vspace{-1.0cm}
\end{figure}

\section{ Learning-based Predictive Beamforming Framework }
Generally, simultaneously optimizing $\bar{\mathbf{\Phi}}_{t}$ and $\mathbf{W}_t$ is intractable due to their non-trivial coupling in (\ref{P1}).
Therefore, we will exploit DL techniques to develop a hierarchical processing approach for addressing this challenging beamforming problem, where we successively handle the outer maximization and the inner maximization in (\ref{P1}) \cite{li2019wirelessly, shen2012distributed}.

As shown in Fig. \ref{Fig:problem_solution}, the developed learning-based beamforming framework consists of two stages: (a) an LoS-based predictive beamforming design and (b) an ICSI-based instantaneous beamforming design.
In stage (a), we aim to determine the optimized LoS-based predictive phase-shift matrix, i.e., $\bar{\mathbf{\Phi}}_t^*$, by solving the outer LoS-based maximization in (\ref{P1}), which can be expressed as
\begin{align}
\max_{\bar{\mathbf{\Phi}}_t, \bar{\mathbf{W}}_t} ~ &\mathbb{E}_{\bar{\mathbf{H}}_{t}|{\mathcal{H}}_t^{\tau}}\left\{
\sum_{k = 1}^K\alpha_k\mathrm{log}_2\left(1 + \frac{|\bar{\mathbf{f}}_{k,t}^H\bar{\mathbf{\Phi}}_t\bar{\mathbf{G}}_t\bar{\mathbf{w}}_{k,t}|^2}{\sum_{j \neq k}^K|\bar{\mathbf{f}}_{k,t}^H\bar{\mathbf{\Phi}}_t\bar{\mathbf{G}}_t\bar{\mathbf{w}}_{j,t}|^2 + \sigma_k^2}\right)\right\} \label{P1_outer} \\
\mathrm{s.t.}\,\,&\sum_{k = 1}^K\|\bar{\mathbf{w}}_{k,t}\|^2\leq P, ~ 0 \leq \bar{\varphi}_{n,t} \leq 2\pi, \forall n \in \mathcal{N}, t \in \mathcal{T}. \notag
\end{align}
Here, $\bar{\mathbf{\Phi}}_t$ is the LoS-based predictive phase-shift matrix for time slot $t$, as defined in (\ref{P1}). $\bar{\mathbf{W}}_t = [\bar{\mathbf{w}}_{1,t},\bar{\mathbf{w}}_{2,t},\cdots,\bar{\mathbf{w}}_{K,t} ] $ is the ICSI-based beamforming matrix for time slot $t$. Note that the aim of stage (a) is to obtain the LoS-based predictive phase-shift matrix and $\bar{\mathbf{W}}_t$ is an auxiliary variable for maximizing (\ref{P1_outer}).
In addition, $\mathbb{E}_{\bar{\mathbf{H}}_{t}|\mathcal{H}_t^{\tau}}\{\cdot\}$ is taken over all the random realizations of $\bar{\mathbf{H}}_{t}$ given the historical LoS channels $\mathcal{H}_t^{\tau}$, as defined in (\ref{P1}).
Note that deriving the objective function of (\ref{P1_outer}) in closed form is in general intractable due to the sophisticated high-dimensional distribution involved in the calculation of the expectation.
As an alternative, a DL approach can be adopted to asymptotically approximate the statistical expectations in (\ref{P1_outer}) by a numerical approach and to optimize performance in a data-driven manner \cite{wang2021interplay}. Given the historical location information of the users, we can calculate the LoS channels for building the training dataset.
Then, using offline training, we can obtain a well-trained DNN to determine the optimized predicted phase-shift matrix $\bar{\mathbf{\Phi}}^*$ based on the historical LoS channels. The details of the DL approach will be given in the next section.
Now, we turn to stage (b).
We first set the phase-shift matrix at the IRS to $\bar{\mathbf{\Phi}}^*$ obtained in stage (a). Then, we employ an existing CE scheme for MISO channels, e.g., \cite{zheng2022survey, liu2022deepresidual, jensen2020optimal}, to obtain the estimated end-to-end ICSI $\hat{\mathbf{h}}_{k,t}$, as defined in (\ref{P1}).
Equipped with $\hat{\mathbf{h}}_{k,t}$, we design the ICSI-based beamforming matrix $\mathbf{W}_t$ to further enhance the WSR of the system.
Thus, the resulting inner maximization problem in (\ref{P1}) can be expressed as
\begin{align}
\max_{{\mathbf{W}}_t} \,\,& \mathbb{E}_{\mathbf{H}_t|\hat{\mathbf{H}}_t} \left\{\sum_{k = 1}^K\alpha_k\mathrm{log}_2\left(1 + \frac{|{\mathbf{h}}_{k,t}^H\mathbf{w}_{k,t}|^2}{{\sum_{j \neq k}^K}|{\mathbf{h}}_{k,t}^H\mathbf{w}_{j,t}|^2 + \sigma_k^2} \right)\right\} \label{P2} \\
\mathrm{s.t.}\,\,&\sum_{k = 1}^K\|\mathbf{w}_{k,t}\|^2\leq P,  t \in \mathcal{T}. \notag
\end{align}
To handle this problem, we also adopt a DL approach to obtain the ICSI-based transmit beamforming matrix $\mathbf{W}_t^*$.
The proposed framework is illustrated in Fig. \ref{Fig:problem_solution}. We tackle the original sophisticated problem by two-stage hierarchical processing and obtain a solution, consisting of the LoS-based phase-shift matrix $\bar{\mathbf{\Phi}}_t^*$ and the ICSI-based transmit beamforming matrix $\mathbf{W}_t^*$.

\textbf{Remark 1}: Note that the proposed predictive approach in Fig. \ref{Fig:problem_solution} can be applied to various design problems for IRS-assisted systems, such as energy efficiency optimization, physical layer security enhancement \cite{wu2019beamforming, guo2020weighted, wang2021interplay, huang2020reconfigurable, gao2020unsupervised}, etc., to facilitate practical implementation with low CE overhead.
Meanwhile, the DNN adopted in Fig. \ref{Fig:problem_solution} can be replaced by any kind of neural network structure, e.g., convolutional neural network (CNN) \cite{liu2019deep}, residual neural network \cite{liu2022deepresidual}, and recurrent neural network \cite{liu2022learning}.
Thus, the developed predictive approach is a versatile general framework for the practical integration of IRSs into wireless networks, striking a balance between the required CE signaling overhead and beamforming performance.

\section{ Deep Learning for Scalable Predictive Beamforming Design }
In this section, we will provide a realization of the proposed framework depicted in Fig. \ref{Fig:problem_solution}. Specifically, an LA-CLGNN is first designed for accomplishing the maximization in (\ref{P1_outer}) and determining the optimized $\bar{\mathbf{\Phi}}_t^*$. Then, an IA-FNN is developed for the maximization in (\ref{P2}) and optimizing $\mathbf{W}_t^*$ for a given $\bar{\mathbf{\Phi}}_t^*$.
Finally, we summarize the proposed DL-based predictive beamforming algorithm.

\begin{figure}[t]
  \centering
  \includegraphics[width=0.75\linewidth]{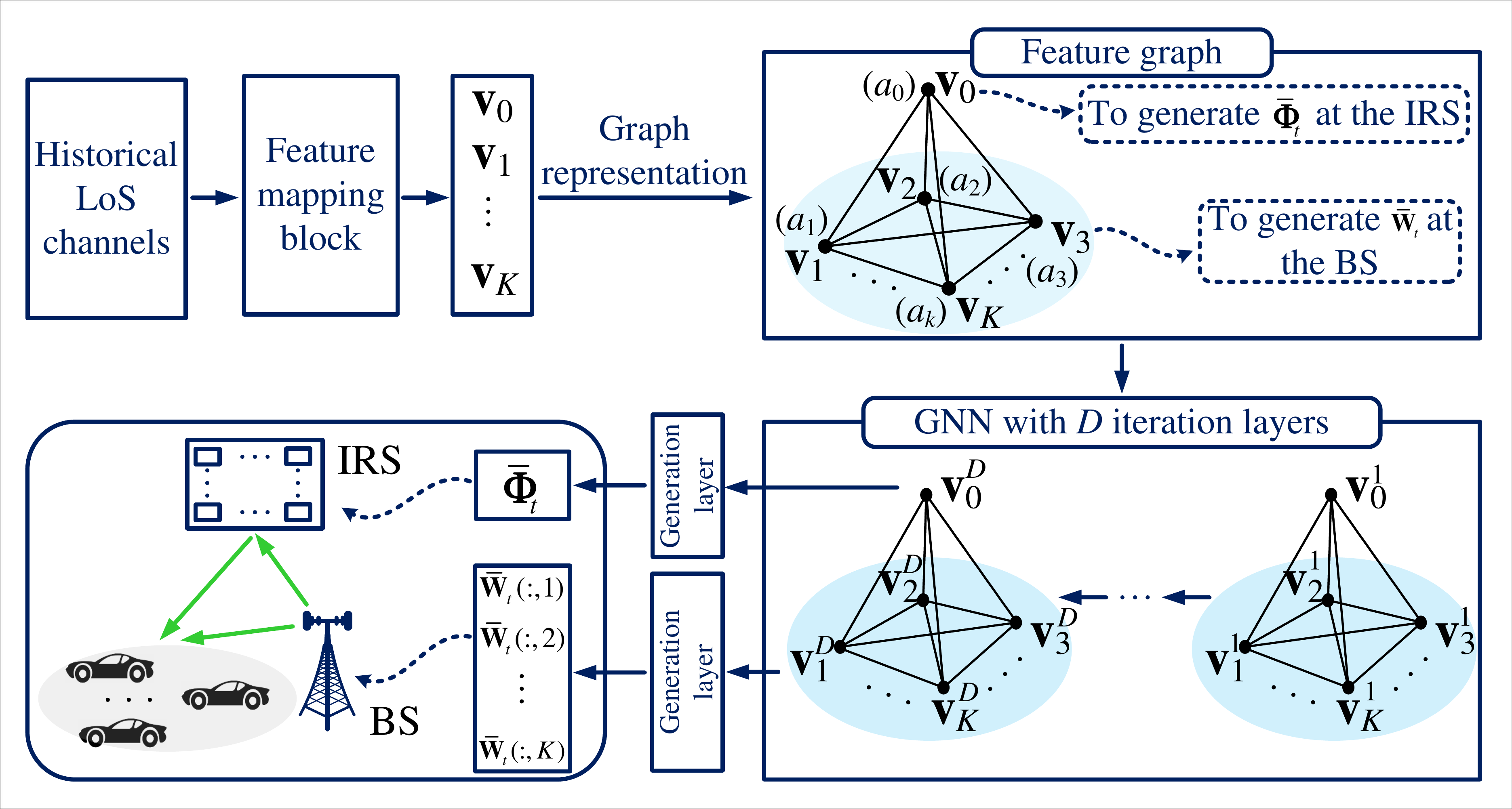}\vspace{-0.5cm}
  \caption{ Illustration of the graph representation in the developed LA-CLGNN. }\label{Fig:graph_representation}
  \vspace{-0.6cm}
\end{figure}

\subsection{ LA-CLGNN for LoS-aware Predictive Beamforming Design }
\begin{figure}[t]
  \centering
  \includegraphics[width=0.9\linewidth]{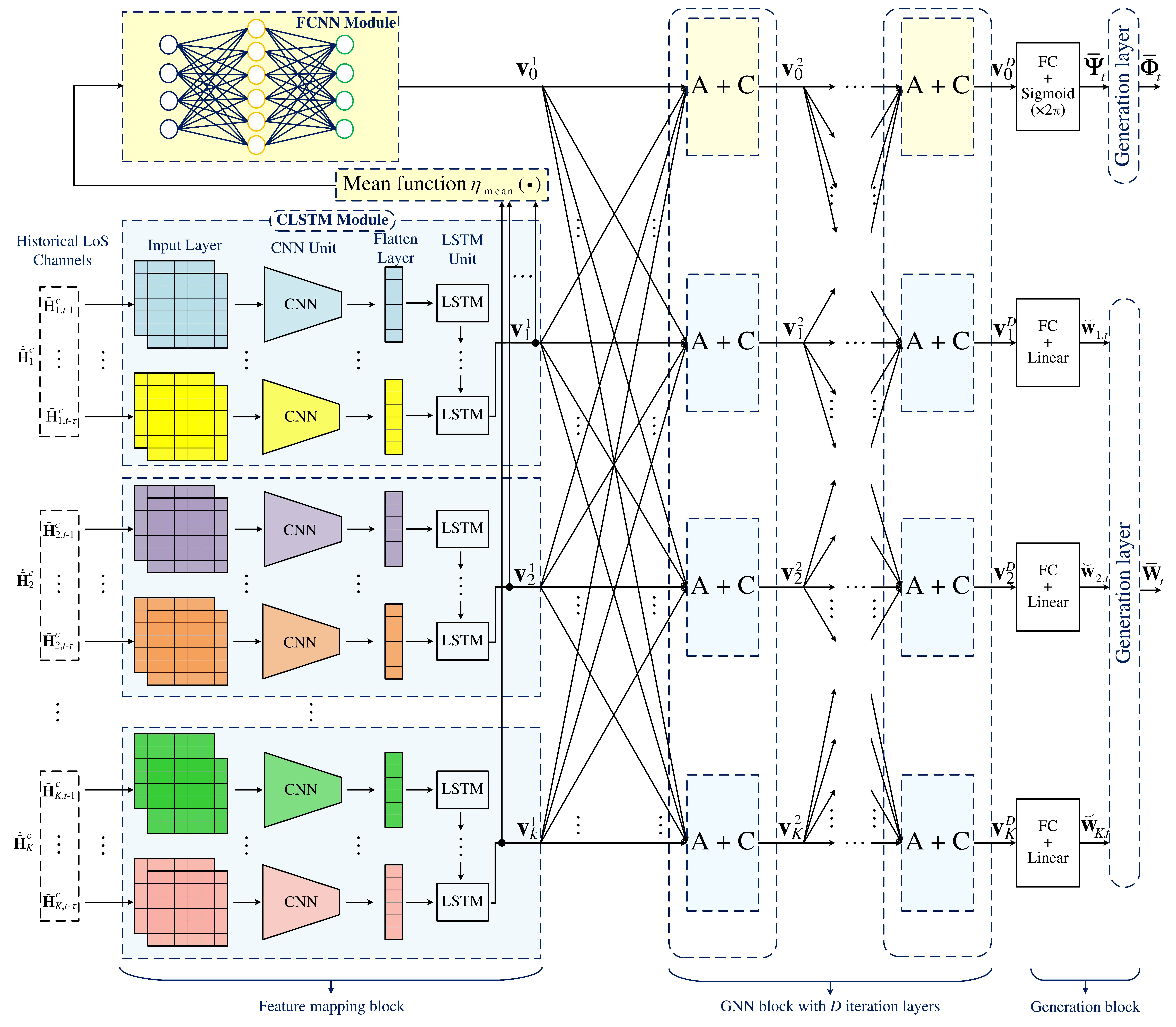}
  \vspace{-0.5cm}
  \caption{ The architecture of the developed LA-CLGNN. }\label{Fig:LA-CLGNN}\vspace{-1.0cm}
\end{figure}

As a realization of the developed framework depicted in Fig. \ref{Fig:problem_solution}, we propose an LA-CLGNN for LoS-based predictive beamforming design to solve the problem in stage (a).
As shown in Fig. \ref{Fig:graph_representation}, we first employ a feature mapping block to exploit the feature vectors from the historical LoS channels for the subsequent neural network processing.
We then represent these feature vectors as a feature graph and adopt a GNN to promote the scalability of the neural network. According to Fig. \ref{Fig:graph_representation}, the feature graph consists of $K+1$ nodes, where $a_k$, $k \in \mathcal{\tilde{K}} \triangleq \{0,1,\cdots,K\}$, denotes the $k$-th node and $\mathbf{v}_k$ is the associated feature vector of $a_k$.
For ease of explanation, node $a_0$ represents the IRS and its associated feature vector $\mathbf{v}_0$ generates $\bar{\mathbf{\Phi}}_t$ for the IRS. Accordingly, nodes $\{a_1,a_2,\cdots,a_K\}$ represent the $K$ users and their associated feature vectors $\{\mathbf{v}_1,\mathbf{v}_2,\cdots,\mathbf{v}_K \}$ produce $\{\bar{\mathbf{W}}_t(:,1),\bar{\mathbf{W}}_t(:,2),\cdots,\bar{\mathbf{W}}_t(:,K) \}$ for generating $\bar{\mathbf{W}}_t$ at the AP.
According to the system model in Section II, the users and the IRS interact with each other, and thus, there exists a link between any two nodes of the feature graph.
The generated feature graph is then sent to a GNN with $D$ iteration layers, where each layer is a fully connected layer. Specifically, the $d$-th layer, $d \in \{1,2,\cdots,D\}$, can be represented as a feature graph with feature vectors $\{\mathbf{v}_0^d,\mathbf{v}_1^d,\cdots,\mathbf{v}_K^d\}$, where $\mathbf{v}_k^d$, $k \in \mathcal{\tilde{K}}$, is the updated feature vector after $d$ updates/iterations.
Given the feature graph, the GNN enables each node to individually extract the features from the other nodes which facilitates the learning of interference management with respect to each node.
More importantly, each node representing a user shares the same neural network structure and parameters, such that it is convenient to add or delete nodes for different system deployments without the need for retraining \cite{shen2020graph, eisen2020optimal, lee2020graph}.
According to the above discussion, in Fig. \ref{Fig:LA-CLGNN}, we provide the detailed LA-CLGNN architecture, which consists of a feature mapping block, a GNN block, and a generation block. We will explain these blocks in the following.


\subsubsection{Feature Mapping Block}
The task of a feature mapping block is to learn the mapping of the input to the feature vectors associated with the desired output.
As shown in Fig. \ref{Fig:LA-CLGNN}, the feature mapping block consists of $K$ CLSTM modules and one fully connected neural network (FCNN) module.
The $k$-th, $k \in \mathcal{K}$, CLSTM module is associated with node $a_k$ and the FCNN module is associated with node $a_0$, as illustrated in Fig. \ref{Fig:graph_representation}.
In particular, each CLSTM module is composed of one input layer, one CNN unit, one flatten layer, and one LSTM unit.
To facilitate the spatial and temporal feature extraction for each user, we stack the $\tau$ historical LoS channel matrices of each user into a supermatrix and use two neural network channels for the real and the imaginary parts of the complex-valued input, respectively. Thus, the input of the $k$-th CLSTM module can be expressed as follows
\begin{equation}\label{Omega}
  \mathbf{\Omega}_{k,t}^{\tau} = \mathcal{F}([ \mathrm{Re}\{\dot{\bar{\mathbf{H}}}^c_{k}\}, \mathrm{Im}\{\dot{\bar{\mathbf{H}}}^c_{k}\} ]) \in \mathbb{R}^{\tau \times N \times M \times 2},
\end{equation}
where
$\dot{\bar{\mathbf{H}}}^c_{k} \triangleq [\bar{\mathbf{H}}_{k, t-1}^c, \cdots, \bar{\mathbf{H}}_{k,t-\tau}^c] \in \mathbb{R}^{N \times \tau M}$ are the historical cascaded LoS channels of $U_k$.
Here, $\mathcal{F}(\cdot)\!\!:\mathbb{R}^{N \times 2\tau M}\mapsto\mathbb{R}^{\tau \times N \times M \times 2}$ is the mapping function and $\bar{\mathbf{H}}_{k,\lambda}^c = \mathrm{diag}(\bar{\mathbf{f}}_{k,\lambda})\bar{\mathbf{G}}$ $ \in \mathbb{C}^{N \times M}$ is the cascaded LoS channel of $U_k$ in time slot $\lambda$.
After the input layer, a CNN unit with rectified linear unit (ReLU) activation function and max pooling operation is adopted to extract the spatial features of the input.
Then, a flatten layer is added after the CNN unit to reshape the feature size for the subsequent LSTM unit.
Finally, the LSTM unit exploits the temporal features by recurrently exploiting the input from the $\tau$ past time slots/steps.
The output of the LSTM in time step $l$, $l \in \{1,2,\cdots,\tau - 1\}$, is the input of the LSTM in time step $l+1$. Specifically, the output of the LSTM in time step $\tau$ is exactly the output of the entire CLSTM module since it exploits the temporal dependencies of the historical channels from all $\tau$ past time slots. Based on this observation, the output of the CLSTM module can be expressed as follows
\begin{equation}\label{}
  \mathbf{v}_k^1 = g_{\mathrm{CL}}(\mathbf{\Omega}_{k,t}^{\tau}), \forall k \in \mathcal{K},
\end{equation}
where $g_{\mathrm{CL}}(\cdot)$ denotes the abstract mathematical model of the CLSTM module.
Since the IRS beamforming design needs to consider the joint effects of all users, for IRS node $a_0$, we adopt an FCNN module, which consists of an input layer, several hidden layers, and an output layer to exploit the features of all user nodes. Without loss of generality, denoting the mathematical model of the FCNN module by $g_{\mathrm{FC}}(\cdot)$, we have
\begin{equation}\label{}
  \mathbf{v}_0^1 = g_{\mathrm{FC}}(\eta_{\mathrm{mean}}(\mathbf{v}_1^1,\mathbf{v}_2^1,\cdots,\mathbf{v}_K^1)),
\end{equation}
with
\begin{equation}\label{mean_function}
  \eta_{\mathrm{mean}}(\mathbf{v}_1^1,\mathbf{v}_2^1,\cdots,\mathbf{v}_K^1) = \frac{1}{K}\sum_{k=1}^K\mathbf{v}_k^1
\end{equation}
being the element-wise mean function for aggregating user features for the overall interference management.
Finally, the obtained feature vectors, i.e., $\{\mathbf{v}_k^1\}_{k\in \tilde{\mathcal{K}}}$, extract the channel features for each node and will be sent to the subsequent iteration block for exploiting the interactions between nodes to further improve the beamforming performance.

\subsubsection{GNN Block}
\begin{figure}[t]
  \centering
  \includegraphics[width=0.7\linewidth]{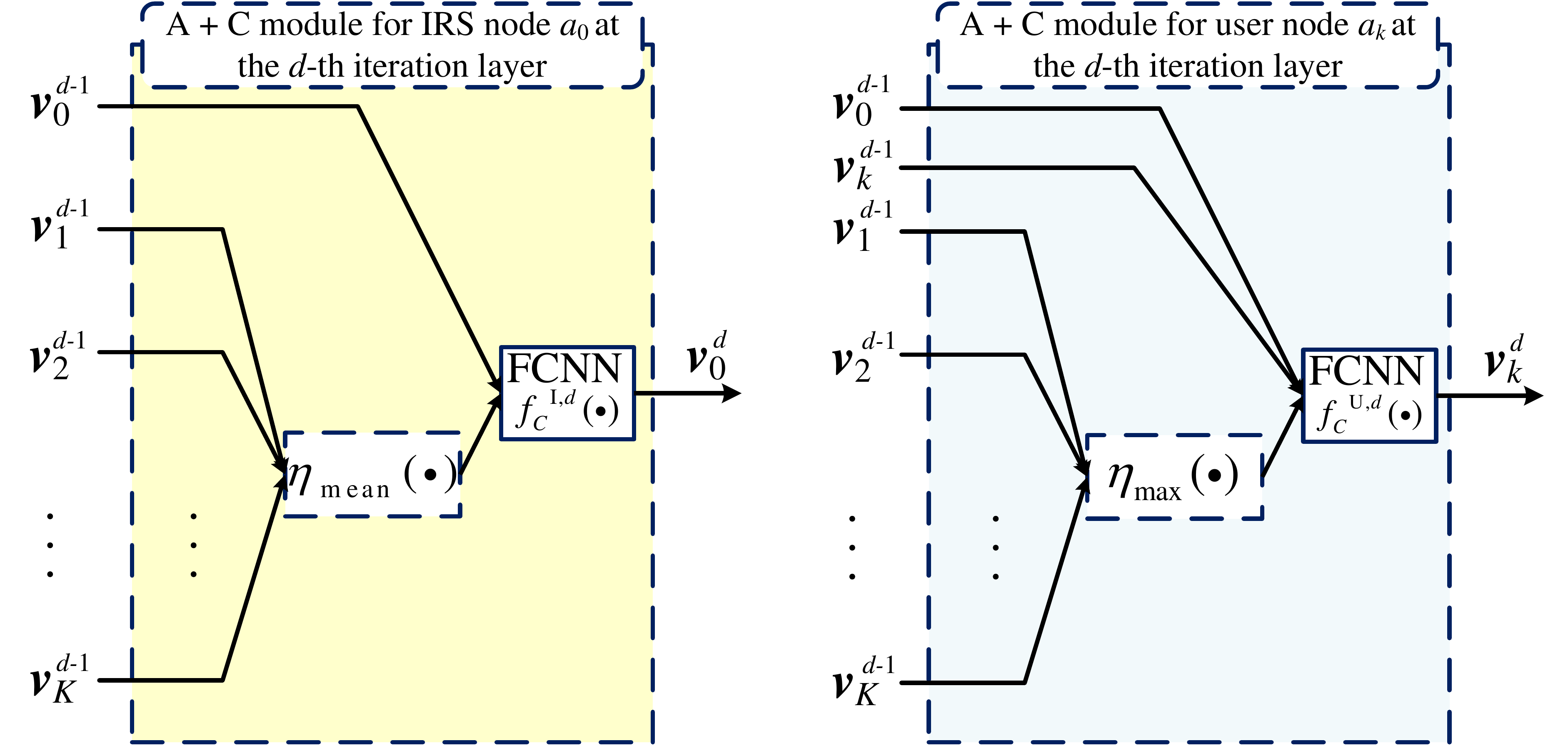}
  \vspace{-0.5cm}
  \caption{ Illustration of the A + C modules in the developed LA-CLGNN. }\label{Fig:A_C}
  \vspace{-1.0cm}
\end{figure}

Given $\{\mathbf{v}_k^1\}_{k\in \tilde{\mathcal{K}}}$, a GNN block equipped with $D$ iteration layers is used to process the features from different nodes for interference management learning.
As illustrated in Fig. \ref{Fig:A_C}, we propose an aggregation-combination module, denoted by ``A + C'' module, for GNN training.
To generate $\mathbf{v}_0^d$ for IRS node $a_0$, we first adopt an element-wise mean function $\eta_{\mathrm{mean}}(\cdot)$ as the aggregation function, as defined in (\ref{mean_function}), to integrate all feature vectors of the user nodes since the IRS needs to serve all the users simultaneously.
Then, we adopt an FCNN to customize the combination function, which yields
\begin{equation}\label{f_C_Id}
  \mathbf{v}_0^d = f_{\mathrm{C}}^{\mathrm{I},d}\left(\mathbf{v}_0^{d-1},
  \eta_{\mathrm{mean}}(\{\mathbf{v}_k^{d-1}\}_{k \in \mathcal{K}} )\right),
\end{equation}
where $f_{\mathrm{C}}^{\mathrm{I},d}(\cdot)$ denotes the combination function for the IRS node at the $d$-th iteration layer.
Similarly, to generate $\mathbf{v}_k^d$ for user node $a_k$, we first employ an element-wise maximization function as the aggregation function to integrate the feature vectors of the other user nodes excluding the IRS node, and then, we adopt an FCNN to customize the combination function to integrate all the feature vectors, i.e.,
\begin{equation}\label{f_C_Ud}
  \mathbf{v}_k^d = f_{\mathrm{C}}^{\mathrm{U},d}\left(\mathbf{v}_0^{d-1}, \mathbf{v}_k^{d-1},
  \eta_{\mathrm{max}}(\{\mathbf{v}_j^{d-1}\}_{j \in \mathcal{K}, j \neq k})\right), \forall k \in \mathcal{K}.
\end{equation}
Here, $f_{\mathrm{C}}^{\mathrm{U},d}(\cdot)$ denotes the combination function for all user nodes in the $d$-th iteration layer.
Since each beamforming vector for each user has to manage the interference from all other users, which is often dominated by the user with the strongest channel, we adopt the element-wise maximization function, i.e.,
\begin{equation}\label{}
  [\eta_{\mathrm{max}}( \mathbf{v}_1,\cdots,\mathbf{v}_K)]_i = \max( [\mathbf{v}_1]_i,\cdots,[\mathbf{v}_K]_i  ).
\end{equation}
As a result, the output of the GNN block with $D$ iteration layers, as shown in Fig. \ref{Fig:LA-CLGNN}, is given by
\begin{equation}\label{}
  \mathbf{v}_0^D = f_{\mathrm{C}}^{\mathrm{I},D}\left(\mathbf{v}_0^{D-1},
  \eta_{\mathrm{mean}}(\{\mathbf{v}_k^{D-1}\}_{k \in \mathcal{K}} )\right)
\end{equation}
and
\begin{equation}\label{}
  \mathbf{v}_k^D = f_{\mathrm{C}}^{\mathrm{U},D}\left(\mathbf{v}_0^{D-1}, \mathbf{v}_k^{D-1},
  \eta_{\mathrm{max}}(\{\mathbf{v}_j^{D-1}\}_{j \in \mathcal{K}, j \neq k} )\right), \forall k \in \mathcal{K}.
\end{equation}

\subsubsection{Generation Block}
After being processed by the GNN block, the computed feature vectors, i.e., $\{\mathbf{v}_k^D\}_{k \in \tilde{\mathcal{K}}}$, are sent to the generation block to obtain the desired phase-shift matrix and transmit beamforming matrix, respectively, cf. Fig. \ref{Fig:LA-CLGNN}.
Specifically, the generation block consists of a fully-connected (FC) layer and a generation layer, where the FC layer is used to crop the feature vectors to a suitable size and the generation layer is then used for the power normalization according to the system requirements.

To generate the phase-shift matrix $\bar{\mathbf{\Phi}}_t$, we first process $\mathbf{v}_0^D$ in the FC layer, resulting in the phase shifts \begin{equation}\label{f_FCI}
  \bar{\mathbf{\Psi}}_t = [\bar{\varphi}_{1,t},\bar{\varphi}_{2,t},\cdots,\bar{\varphi}_{N,t}]^T
  = f_\mathrm{FC}^{\mathrm{I}}(\mathbf{v}_0^D) \in \mathbb{R}^{N\times 1},
\end{equation}
where $f_\mathrm{FC}^{\mathrm{I}}(\cdot)$ represents the FC layer with sigmoid activation function and multiplication with $2\pi$ to satisfy the phase constraint in (\ref{P1_outer}).
Finally, the phase-shift matrix is obtained by a generation layer, which is expressed as
\begin{equation}\label{}
  \bar{\mathbf{\Phi}}_t = \mathrm{diag}\left([e^{j[\bar{\mathbf{\Psi}}_t]_1},
  e^{j[\bar{\mathbf{\Psi}}_t]_2},\cdots,e^{j[\bar{\mathbf{\Psi}}_t]_N}]\right) \in \mathbb{C}^{N\times N}.
\end{equation}
Similarly, to generate transmit beamforming matrix $\bar{\mathbf{W}}_t$, we first process $\mathbf{v}_k^D$, $k \in \mathcal{K}$, in a FC layer with linear activation function, denoted by $f_\mathrm{FC}^{\mathrm{U}}(\cdot)$, to obtain an output vector, i.e.,
\begin{equation}\label{f_FCU}
  \breve{\mathbf{w}}_{k,t}
  = f_\mathrm{FC}^{\mathrm{U}}(\mathbf{v}_k^D) \in \mathbb{R}^{2M\times 1},
\end{equation}
where $\breve{\mathbf{w}}_{k,t}(1:M)$ and $\breve{\mathbf{w}}_{k,t}(M+1:2M)$ correspond to the real and imaginary parts of the complex-valued transmit beamforming vector, respectively.
Considering the power constraint in (\ref{P1_outer}), a generation layer is added after the FC layer for normalization, i.e.,
\begin{equation}\label{Z_t}
  \mathbf{Z}_t = \sqrt{P}\frac{\breve{\mathbf{W}}_t}{\|\breve{\mathbf{W}}_t\|_F}
  \in \mathbb{C}^{2M\times K},
\end{equation}
where $\breve{\mathbf{W}}_t =[\breve{\mathbf{w}}_{1,t},\breve{\mathbf{w}}_{2,t},\cdots,\breve{\mathbf{w}}_{K,t}]$.
Finally, the normalized transmit beamforming matrix can be expressed as
\begin{equation}\label{W_t_generation}
  \bar{\mathbf{W}}_t = [\mathbf{Z}_t(1:M,:) + j\mathbf{Z}_t(M+1:2M,:)]
  \in \mathbb{C}^{M\times K},
\end{equation}
where $j=\sqrt{-1}$ denotes the imaginary unit and $\mathbf{Z}_t(m_1:m_2,:)$ denotes the submatrix of $\mathbf{Z}_t$ ranging from the $m_1$-th row to the $m_2$-th row of $\mathbf{Z}_t$.

\textbf{Remark 2}: We note that for all user nodes $\{a_k\}_{k\in \mathcal{K}}$, we adopt identical $g_{\mathrm{CL}}(\cdot)$, $f_{\mathrm{C}}^{\mathrm{I},d}(\cdot)$, $f_{\mathrm{C}}^{\mathrm{U},d}(\cdot)$, and $f_{\mathrm{FC}}^{\mathrm{U}}(\cdot)$ for all modules/blocks/layers to update the feature vectors when generating the dedicated transmit beamforming vector for each user.
In this case, as will be verified in Section IV, when the number of users varies, a well-trained LA-CLGNN only needs to update the number of nodes in the GNN accordingly without the need for additional neural network training.
This is not possible with other existing DL-based approaches, e.g., \cite{wang2021interplay, huang2020reconfigurable, gao2020unsupervised}.
Table \ref{Tab:Hyperparameters LA-CLNet} provides a possible hyperparameter setting for the LA-CLGNN.

\begin{table}[t]
\normalsize
\caption{Hyperparameters of LA-CLGNN}\label{Tab:Hyperparameters LA-CLNet}
\centering
\small
\renewcommand{\arraystretch}{0.95}
\begin{tabular}{c c c}
  \hline
   \multicolumn{3}{l}{\textbf{Input}: $\mathbf{\Omega}_{1,t}^{\tau},\mathbf{\Omega}_{2,t}^{\tau},\cdots,
   \mathbf{\Omega}_{K,t}^{\tau}$ where each $\mathbf{\Omega}_{k,t}^{\tau} \in \mathbb{R}^{\tau \times N \times M \times 2}$ }  \\
  \hline
  \multicolumn{3}{l}{\textbf{Feature Mapping Block}: }  \\
  \hspace{0.1cm} \textbf{Names} & \textbf{Parameters} &  \hspace{0.3cm} \textbf{Values}   \\
  \hspace{0.1cm} FCNN module & Size of each layer & \hspace{0.3cm}  $ 4 \times 3 \times 3 \times 2$   \\
  \hspace{0.1cm} CLSTM module - CNN Unit & Size of convolutional filters & \hspace{0.3cm}  $ 4 \times 3 \times 3 \times 2$   \\
  \hspace{0.1cm} CLSTM module - CNN Unit & Activation function & \hspace{0.3cm}  ReLU  \\
  \hspace{0.1cm} CLSTM module - CNN Unit & Size of pooling filters & \hspace{0.3cm}  $ 3 \times 3$   \\
  \hspace{0.1cm} CLSTM module - Flatten Layer & Size of layer & \hspace{0.3cm} $64 \times 1$ \\
  \hspace{0.1cm} CLSTM module - LSTM Unit & Size of output  & \hspace{0.3cm} $64 \times 1$    \\

  \multicolumn{3}{l}{\textbf{GNN Block}: }  \\
  \hspace{0.1cm} \textbf{Names} & \textbf{Parameters} &  \hspace{0.3cm} \textbf{Values}   \\
  \hspace{0.1cm}$f_C^{\mathrm{I},d}(\cdot) $ in (\ref{f_C_Id}) & Size and activation function  & \hspace{0.3cm}  $ 512 \times 512 \times 512 $ and ReLU   \\
  \hspace{0.1cm}$f_C^{\mathrm{U},d}(\cdot) $ in (\ref{f_C_Ud}) & Size and activation function  & \hspace{0.3cm}  $ 512 \times 512 \times 512 $ and ReLU   \\
  \multicolumn{3}{l}{\textbf{Generation Block}: }  \\
  \hspace{0.1cm} \textbf{Names} & \textbf{Parameters} &  \hspace{0.3cm} \textbf{Values}   \\
  \hspace{0.1cm}$f_{\mathrm{FC}}^{\mathrm{I}}(\cdot)$ in (\ref{f_FCI}) & Size and activation function  & \hspace{0.3cm}  $ 512 \times 512 \times 512 $ and Sigmoid   \\
  \hspace{0.1cm}$f_{\mathrm{FC}}^{\mathrm{U}}(\cdot)$ in (\ref{f_FCU}) & Size and activation function  & \hspace{0.3cm}  $ 512 \times 512 \times 512 $ and Linear  \\
  \hline
   \multicolumn{3}{l}{\textbf{Output}: $\bar{\mathbf{\Phi}}_t \in \mathbb{C}^{N\times N}$ and $\bar{\mathbf{W}}_t \in \mathbb{C}^{M \times K}$ } \\
  \hline
\end{tabular}
\vspace{-0.8cm}
\end{table}

\subsubsection{LA-CLGNN Training}
Although obtaining the optimal solution to (\ref{P1_outer}) is intractable, we can exploit unsupervised offline training to update the neural network parameters in a data-driven manner.
Given an unlabeled training set:
\begin{equation}\label{}
  \mathcal{X} = \{ ( \mathbf{\Omega}_t^{\tau(1)},\bar{\mathbf{H}}_{t}^{ (1)} ),  \cdots, ( \mathbf{\Omega}_t^{\tau(N_t)},\bar{\mathbf{H}}_{t}^{ (N_t)} )  \}.
\end{equation}
Here, $( \mathbf{\Omega}_t^{\tau(i)},\bar{\mathbf{H}}_{t}^{ (i)} )$ is the $i$-th training example, $i \in \{1,2,\cdots,N_t\}$, of $\mathcal{X}$ with historical LoS channel input $\mathbf{\Omega}_t^{\tau(i)} = \{\mathbf{\Omega}_{1,t}^{\tau(i)}, \mathbf{\Omega}_{2,t}^{\tau(i)},\cdots,\mathbf{\Omega}_{K,t}^{\tau(i)}\}$, where $\bar{\mathbf{H}}_{t}^{ (i)}$ and $\mathbf{\Omega}_{k,t}^{\tau(i)}$ are defined in (\ref{P1}) and (\ref{Omega}), respectively.
Based on the problem formulation in (\ref{P1_outer}), the cost function of the LA-CLGNN can be expressed as
\begin{equation}\label{J_LA-CLGNN}
  J_{\mathrm{LA-CLGNN}}(\omega) = -\frac{1}{N_t}\sum_{i=1}^{N_t} \sum_{k = 1}^K\alpha_k\mathrm{log}_2(1 + \bar{\gamma}_{k,t}^{(i)}({\omega}))
\end{equation}
with
\begin{equation}\label{}
\bar{\gamma}_{k,t}^{(i)}({\omega}) = \frac{|(\bar{\mathbf{f}}_{k,t}^{(i)})^H\bar{\mathbf{\Phi}}_t^{(i)}(\omega)\bar{\mathbf{G}}_t^{(i)}\bar{\mathbf{w}}_{k,t}^{(i)}(\omega)|^2}
{{\sum_{j \neq k}^K}|(\bar{\mathbf{f}}_{k,t}^{(i)})^H\bar{\mathbf{\Phi}}_t^{(i)}(\omega)\bar{\mathbf{G}}_t^{(i)}\bar{\mathbf{w}}_{j,t}^{(i)}(\omega)|^2 + \sigma_k^2}.
\end{equation}
Here, $\bar{\mathbf{f}}_t^{(i)}$ and $\bar{\mathbf{G}}_t^{(i)}$ are the LoS channels at time slot $t$.
$\bar{\mathbf{\Phi}}_t^{(i)}(\omega)$ and $\bar{\mathbf{w}}_{k,t}^{(i)}(\omega)$ denote the output of the LA-CLGNN employing the network parameters $\omega$, based on input $\mathbf{\Omega}_t^{\tau(i)}$.
After offline training via the backpropagation algorithm, we can obtain the well-trained LA-CLGNN, i.e.,
\begin{equation}\label{Psi*}
  [\bar{\mathbf{\Phi}}_t^*,\bar{\mathbf{W}}_t^*]= \vartheta_{\omega^*}(\mathbf{\Omega}_{t}^{\tau}).
\end{equation}
Here, $\bar{\mathbf{\Phi}}_t^*$ and $\bar{\mathbf{W}}_t^*$ are the well-trained phase shifts and beamforming matrix, respectively.
$\vartheta_{\omega^*}(\cdot)$ represents the entire LA-CLGNN with the well-trained network parameters $\omega^*$.
We note that $\bar{\mathbf{\Phi}}_t^*$ is adopted as the IRS phase-shift matrix in the next time slot, while $\bar{\mathbf{W}}_t^*$ is only an auxiliary parameter obtained in the course of optimization.

\subsection{ IA-FNN for ICSI-based Instantaneous Beamforming Design }
\begin{figure}[t]
  \centering
  \includegraphics[width=0.6\linewidth]{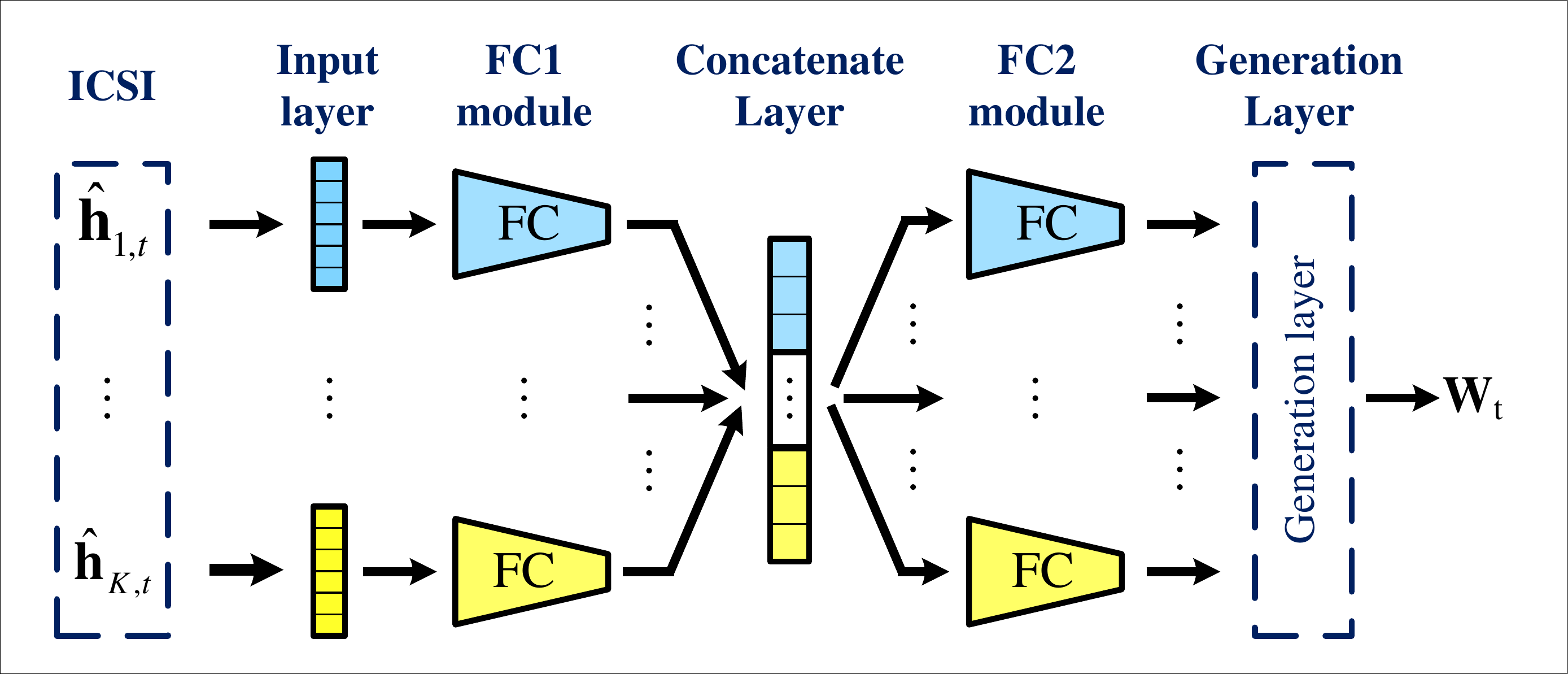}\vspace{-0.2cm}
  \caption{ The architecture of the developed IA-FNN. }\label{Fig:IA-FNN}\vspace{-0.5cm}
\end{figure}

Given $\bar{\mathbf{\Phi}}_t^*$ from the well-trained LA-CLGNN, the ICSI of each AP-IRS-$U_k$ link can be estimated, i.e., $\hat{\mathbf{H}}_t$ as defined in (\ref{P2}).
In the following, we propose an IA-FNN to determine the ICSI-based instantaneous beamforming design, as illustrated in stage (b) of Fig. \ref{Fig:problem_solution}.
The proposed IA-FNN is a multiple-input multiple-output neural network architecture, as shown in Fig. \ref{Fig:IA-FNN}, which consists of an input layer, an FC1 module, a concatenate layer, an FC2 module, and a generation layer.
The hyperparameters of IA-FNN are provided in Table \ref{Tab:Hyperparameters IA-FNN}.
In particular, the FC1 and FC2 modules are FCNNs, where the last layer has a linear activation function and the other layers have ReLU activation functions.
Similar to the LA-CLGNN, the input of IA-FNN can be expressed as $\ddot{\mathbf{H}}_{t} = \mathcal{G}([ \mathrm{Re}\{\hat{\mathbf{H}}_{t}\}, \mathrm{Im}\{\hat{\mathbf{H}}_{t}\} ])$, where $\mathcal{G}(\cdot)\!\!:\mathbb{R}^{M \times 2K}\mapsto\mathbb{R}^{M \times K \times 2}$ is a mapping function.
After the processing of the hidden layers, we obtain the output vectors from the FC2 module and they are forwarded to the generation layer.
Similar to (\ref{f_FCU})-(\ref{W_t_generation}), the generation layer finally outputs beamforming matrix ${\mathbf{W}}_t = f_{\varsigma}^{\mathrm{IF}} (\ddot{\mathbf{H}}_t) \in \mathbb{C}^{M \times K}$, where $f_{\varsigma}^{\mathrm{IF}}(\cdot)$ is the mathematical representation of IA-FNN with neural network parameters  $\varsigma$.

\begin{table}[t]
\normalsize
\caption{Hyperparameters of the proposed IA-FNN}\vspace{-0.2cm}\label{Tab:Hyperparameters IA-FNN}
\centering
\small
\renewcommand{\arraystretch}{1.15}
\begin{tabular}{c c c}
  \hline
   \multicolumn{3}{l}{\textbf{Input}: $\ddot{\mathbf{H}}_t$ with the size of $M \times K \times 2$}  \\
  \hline
  \hspace{0.1cm} \textbf{Layers/Modules} & \textbf{Parameters} &  \hspace{0.3cm} \textbf{Default Values}   \\
  \hspace{0.1cm} FC1 module & Sizes of weights & \hspace{0.3cm}  $32 \times 16 \times 16 $   \\
  Concatenate layer & Size of output & \hspace{0.3cm}  $ K \times 16 $   \\
  FC2 module & Sizes of weights & \hspace{0.3cm}  $64 \times 32 \times 12$   \\
  \hspace{0.1cm} Generation layer & Size of output & \hspace{0.3cm}  $ M \times K$   \\
  \hline
   \multicolumn{3}{l}{\textbf{Output}: $\mathbf{W}_t \in \mathbb{C}^{M \times K}$ }    \\
  \hline
\end{tabular}
\vspace{-0.6cm}
\end{table}

To improve the robustness of the beamforming design against ICSI imperfection, we adopt an offline training scheme for IA-FNN.
Let $\tilde{\mathcal{X}} = \{ (\ddot{\mathbf{H}}_{t}^{ (1)},\mathbf{H}_{t}^{ (1)} ),  \cdots, (\ddot{\mathbf{H}}_{t}^{ (\tilde{N}_t)},\mathbf{H}_{t}^{ (\tilde{N}_t)} ) \}$ denote a given training dataset, where $(\ddot{\mathbf{H}}_{t}^{ (i)},\mathbf{H}_{t}^{(i)})$ is the $i$-th training example, $i\in\{1,2,\cdots,\tilde{N}_t\}$, of $\tilde{\mathcal{X}}$ with $\ddot{\mathbf{H}}_{t}^{ (i)}$ and $\mathbf{H}_{t}^{(i)}$ being the estimated ICSI and the perfect ICSI, respectively.
According to the problem formulated in (\ref{P2}), the cost function of IA-FNN is given by
\begin{equation}\label{J_IA-FNN}
  J_{\mathrm{IA-FNN}}(\varsigma) = -\frac{1}{\tilde{N}_t}\sum_{i=1}^{\tilde{N}_t}\sum_{k = 1}^K\alpha_k\mathrm{log}_2 \left(1 + \frac{|(\mathbf{h}_{k,t}^{(i)})^H\mathbf{w}_{k,t}^{(i)}(\varsigma)|^2}{{\sum_{j \neq k}^K}|(\mathbf{h}_{k,t}^{(i)})^H\mathbf{w}_{j,t}^{(i)}(\varsigma)|^2 + \sigma_k^2}  \right),
\end{equation}
where $\mathbf{w}_{k,t}^{(i)}(\varsigma)$ denotes the output of IA-FNN with parameter $\varsigma$, based on input $\ddot{\mathbf{H}}_t^{(i)}$.
After offline training via the backpropagation algorithm, we obtain the well-trained IA-FNN.
Thus, the well-trained beamforming matrix ${\mathbf{W}}_t^*$ can be expressed as
\begin{equation}\label{}
  \mathbf{W}_t^* = f_{\varsigma^*}^{\mathrm{IF}} (\ddot{\mathbf{H}}_t),
\end{equation}
where $\varsigma^*$ denotes the well-trained network parameters.


\begin{table}[t]
\small
\centering
\begin{tabular}{l}
\toprule[1.8pt] \vspace{-0.8cm}\\
\hspace{-0.1cm} \textbf{Algorithm 1} { DL-based Predictive Beamforming (DLPB) Algorithm } \\
\toprule[1.8pt] \vspace{-0.6cm}\\
\textbf{Offline Training:} \\
1:\hspace{0.25cm}\textbf{Input:} Training sets $\mathcal{X}$ and $\tilde{\mathcal{X}}$ for LA-CLGNN and IA-FNN, respectively  \\
2:\hspace{0.3cm}\textbf{do} neural network training via backpropagation algorithm \\
3:\hspace{0.65cm} Update $\omega$ and $\varsigma$ by minimizing $J_{\mathrm{LA-CLGNN}}(\omega)$ in (\ref{J_LA-CLGNN}) and $J_{\mathrm{IA-FNN}}(\varsigma)$ in (\ref{J_IA-FNN}), respectively \\
5:\hspace{0.25cm}\textbf{Output}:  Well-trained $f_{\omega^*}^{\mathrm{LC}}(\cdot)$ and  $f_{\varsigma^*}^{\mathrm{IF}}(\cdot)$ \\
\textbf{Online Prediction:} LoS-based Predictive Beamforming \\
6:\hspace{0.25cm}\textbf{Input:} Test data $\mathbf{\Omega}_{t}^{\tau} $  \\
7:\hspace{0.8cm}Compute the predictive beamforming matrix with the well-trained LA-CLGNN  \\
8:\hspace{0.25cm}\textbf{Output:} $[\bar{\mathbf{\Phi}}_t^*,\bar{\mathbf{W}}_t^*]= f_{\omega^*}^{\mathrm{LC}}(\mathbf{\Omega}_{t}^{\tau}))$ \\
\textbf{Online Beamforming:} ICSI-based Instantaneous Beamforming \\
9:\hspace{0.25cm}\textbf{Input:} Test data $\ddot{\mathbf{H}}_{t} $ for IA-FNN obtained by MISO CE for given $\mathbf{\Phi}_t = \bar{\mathbf{\Phi}}_t^*$  \\
10:\hspace{0.56cm} Compute beamforming matrix with the well-trained IA-FNN \\
11:\hspace{0.1cm}\textbf{Output:} $\mathbf{W}_t^* = f_{\varsigma^*}^{\mathrm{IF}} (\ddot{\mathbf{H}}_t)$ \\
\bottomrule[1.8pt]
\end{tabular}
\vspace{-0.8cm}
\end{table}

\subsection{ DL-based Predictive Beamforming Algorithm }
Based on the above discussions, the proposed DL-based predictive beamforming (DLPB) algorithm is summarized in Algorithm 1, which involves offline training, online prediction, and online beamforming.

\section{ Numerical Results }
In this section, we focus on a typical IRS-MUC system and provide extensive simulation results to evaluate the performance of the proposed framework.
In the considered IRS-MUC system, there is one $N = N_y \times N_z$-element IRS supporting the communications between one AP equipped with an $M$-element antenna array and $K$ single-antenna users.

\subsection{Simulation Setting}
Unless specified otherwise, we set $M = 6$, $N=100$ with $ N_y = N_z = 10 $, $K = 3$, and $\alpha_k=1$, $\forall k \in \mathcal{K}$.
The distances between two adjacent antenna/IRS elements are set as $\Delta d_\mathrm{A} = \Delta d_{\mathrm{I}y} = \Delta d_{\mathrm{I}z} = \frac{1}{2}\lambda_c$, respectively.
The locations of AP and IRS are set as $\mathbf{L}_\mathrm{A} = [2,0,20]^T~\mathrm{m}$ and $\mathbf{L}_\mathrm{I} = [0,50,25]^T~\mathrm{m}$, respectively, in Fig. \ref{Fig:scenario}.
Without loss of generality, the users' initial locations are randomly distributed within a rectangular service area, where the coordinates of the top-left corner and the bottom-right corner are $(3,50,0)$ and $(6,60,0)$, respectively.
According to the movement function defined in (\ref{movement_function}), we set $A_1 = 8~\mathrm{m}/\mathrm{s}$ (around $30~\mathrm{km}/\mathrm{h}$), $A_2 = 10~\mathrm{m}/\mathrm{s}$ (around $40~\mathrm{km}/\mathrm{h}$), $B_1 = -\pi/18~\mathrm{rad}$ (i.e., $-10^{\circ}$), $B_2 = \pi/18~\mathrm{rad}$ (i.e., $10^{\circ}$), $\Delta T = 0.02$ s, and $\sigma_u =0.1$ \cite{9076668}.
In addition, the path losses in this paper are modeled as
$\alpha^{\mathrm{AI}} = \beta_0(d^{\mathrm{AI}}/D_0)^{-\eta_{\mathrm{AI}}}$ and $\alpha^{\mathrm{IU}}_{k,t} = \beta_0(d^\mathrm{IU}_{k,t}/D_0)^{-\eta_{k}}$, $\forall k$, where $\beta_0 = -30~\mathrm{dB}$ represents the path loss at reference distance $D_0 = 1~\mathrm{m}$, and $\eta_{\mathrm{AI}} = 2.2$ and $\eta_k = 2.8$ are the path loss exponents of the AP-IRS and IRS-$U_k$ links, respectively.
The noise variances are set as $\sigma_k^2 = -80~\mathrm{dBm}$, $\forall k$.
In addition, we assume that the CE error vector $\Delta \mathbf{h}$ is a CSCG random vector\cite{guo2020weighted}.
Each element in $\Delta \mathbf{h}$ has zero mean and is generated based on the same normalized mean square error (NMSE)\footnotemark\footnotetext{The NMSE is independent of the velocity of the users since the user location is assumed to be constant in each time slot, see Section II.}: $\varrho = \frac{\mathbb{E}[\|\Delta \mathbf{h}\|^2]}{\mathbb{E}[\|\mathbf{h}_{k,t}\|^2]}$.
To evaluate the performance of the proposed framework, we assume that the proposed DLPB method is conducted under imperfect ICSI with $\varrho=0.1$ \cite{liu2022deepresidual}, while for the benchmark schemes, we assume perfect ICSI.
For the DLPB scheme proposed in Algorithm 1, unless otherwise specified, we set $\tau = 5$ and $N_t = \tilde{N}_t = 2,000$. Also, the hyperparameters of LA-CLGNN and IA-FNN are set according to Tables \ref{Tab:Hyperparameters LA-CLNet} and \ref{Tab:Hyperparameters IA-FNN}, respectively.

Furthermore, to evaluate the performance of the proposed scheme, three benchmark schemes are considered for comparison:
\begin{itemize}
  \item Benchmark 1 (FP-ICSI): An FP-based method \cite{guo2020weighted} is adopted to maximize the average system WSR for perfect full ICSI. In particular, this iterative algorithm is considered to be converged when the WSR increment between two consecutive iterations does not exceed $10^{-2}$.

  \item Benchmark 2 (Naive FP): In this scheme, only the perfect full ICSI at time slot $t-\tau_0$, $\tau_0 = 5$, is available, i.e., $\mathbf{H}_{t-\tau_0}$. Then, the FP-based method \cite{guo2020weighted} is adopted and it naively treats $\mathbf{H}_{t-\tau_0}$ as the actual ICSI in time slot $t$ for beamforming design.

  \item Benchmark 3 (Random PS): Random phase shifts (PS) are adopted and maximum-ratio transmission (MRT) \cite{tse2005fundamentals} is used for beamforming.
\end{itemize}

In addition, all simulation results were generated by averaging over $2,000$ Monte Carlo realizations.

\begin{figure}[t]
  \centering
  \includegraphics[width=3.2in,height=2.8in]{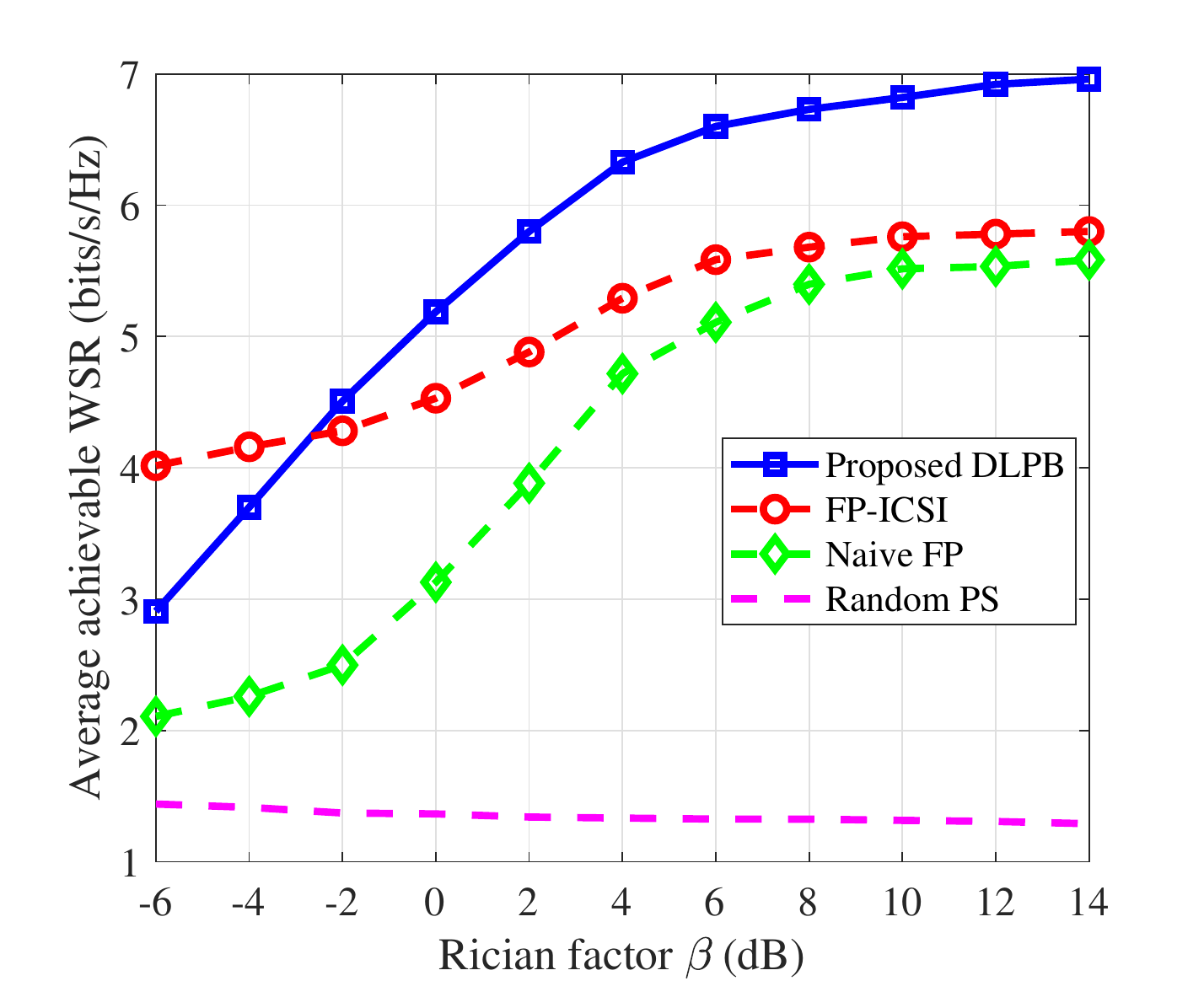}\vspace{-0.5cm}
  \caption{ Average achievable WSR versus $\beta$ for $P = 30~\mathrm{dBm}$.
  }\label{Fig:sum_rate_beta}\vspace{-0.5cm}
\end{figure}

\subsection{WSR Performance}
Fig. \ref{Fig:sum_rate_beta} depicts the average achievable WSR versus the Rician factor, i.e., $\beta^{\mathrm{AI}}$ and $\beta^{\mathrm{IU}}_{k,t}$ defined in (\ref{G_model}) and (\ref{f_model}), respectively.
For simplicity of presentation, we set $\beta^{\mathrm{AI}} = \beta^{\mathrm{IU}}_{k,t} = \beta, \forall k,t$.
As can be observed, the WSR for FP-ICSI, the proposed DLPB scheme, and naive FP increases with $\beta$.
This is because for larger $\beta$, the channels of the AP-IRS link become more correlated and predictable, which is beneficial for improving the beamforming gain.
Furthermore, when $\beta$ is sufficiently large, all channels are dominated by their LoS components and the impact of the random NLoS components is minimal. Thus, the WSRs of all considered algorithms tend to converge to a constant value, as indicated in Fig. \ref{Fig:sum_rate_beta}.
In particular, it is observed that the proposed DLPB scheme outperforms the FP-ICSI method for $\beta \geq -2~\mathrm{dB}$, achieving the best performance among all considered schemes.
The reason for this is that when $\beta$ becomes non-negligible, the deterministic LoS components of the AP-IRS-user links start playing an important role, which is beneficial for the proposed predictive beam alignment as it exploits the historical information of the LoS channels for determining the IRS phase shifts.
On the other hand, the FP algorithm used in FP-ICSI is a sub-optimal method which has limited optimization performance, while the proposed DLPB can exploit the non-linear units of the employed powerful neural networks to improve the WSR performance.
Furthermore, the proposed method can achieve a significant performance gain compared to the naive FP method and the random PS method.
This is expected since the naive FP method only exploits outdated ICSI for beamforming design and random PS can not take advantage of the IRS as random passive beamforming does not always align with the user channels leading to an underutilization of the system resources.
In contrast, the proposed DLBP scheme is designed to make full use of the historical LoS channels and the end-to-end ICSI for beamforming design.

\begin{figure}
\centering
\subfigure[Average WSR versus $P$ with $\beta = 2~\mathrm{dB}$.]{
\includegraphics[width=7.7cm,height=7.0cm]{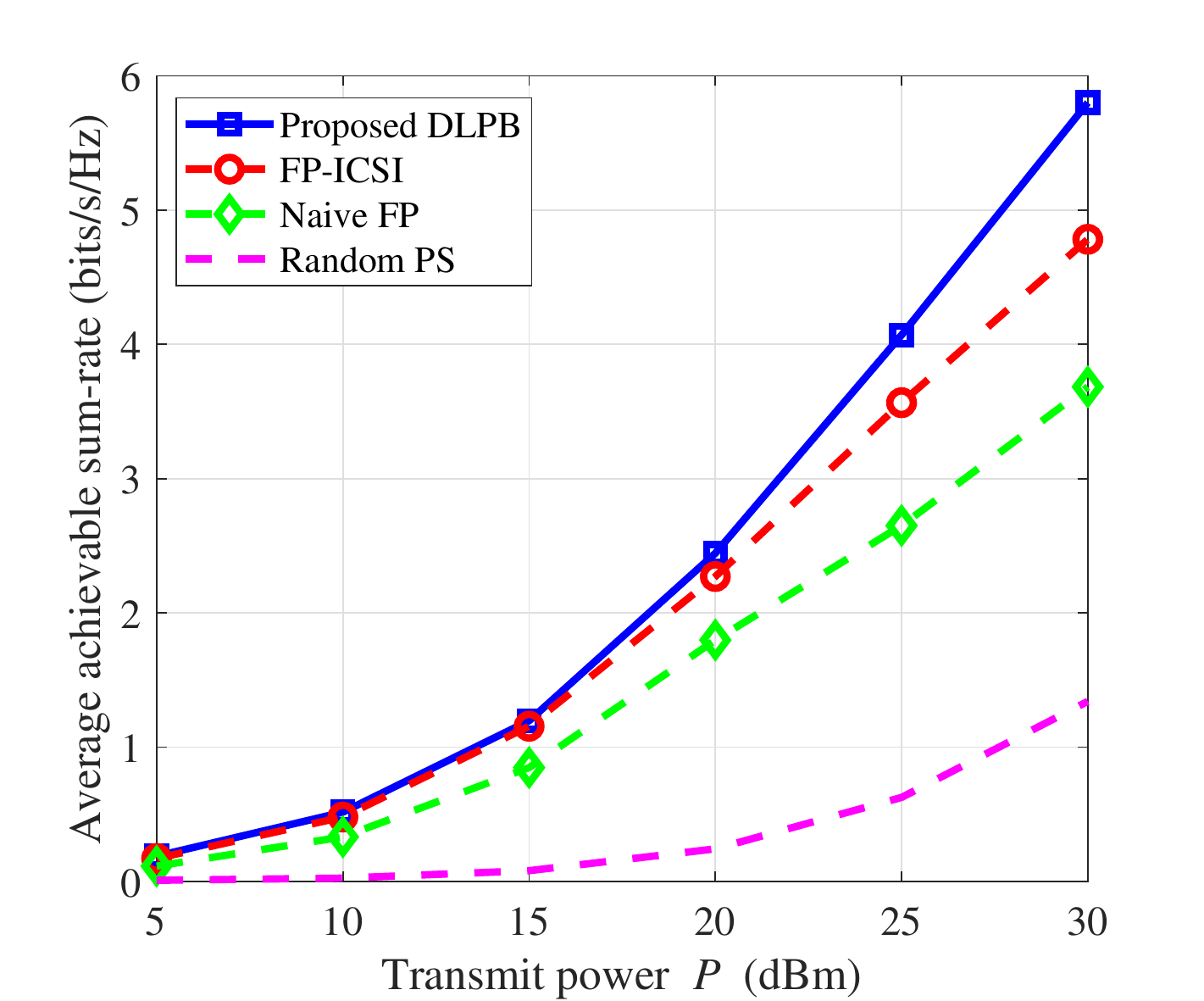}\label{Fig:sum_rate_power}
}
\subfigure[Time-average WSR versus $P$ with $\beta = 2~\mathrm{dB}$ taking into account the CE overhead.]{
\includegraphics[width=7.7cm,height=7.0cm]{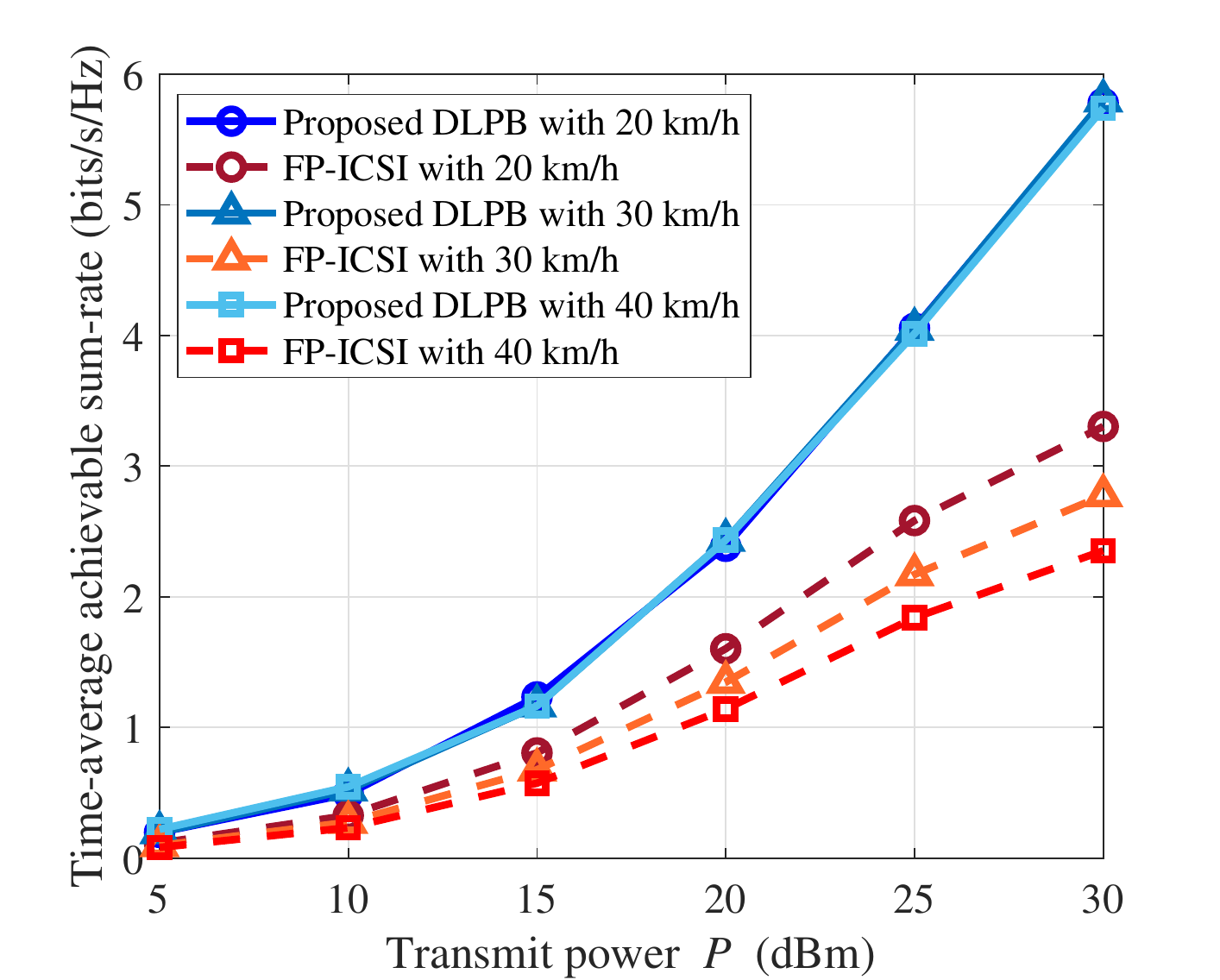}\label{Fig:sum_rate_power_protocol}
}
\DeclareGraphicsExtensions.
\caption{System WSR performance versus the maximum transmit power for $\beta = 2~\mathrm{dB}$.}
\label{Fig:rate_power}\vspace{-0.8cm}
\end{figure}

In Fig. \ref{Fig:rate_power}, we show simulation results for the system WSR versus the transmit power, where both the average achievable WSR and the protocol-based time-average achievable WSR are considered.
Fig. \ref{Fig:sum_rate_power} reveals that the proposed method achieves the best performance among all considered schemes, which is in line with the results in Fig. \ref{Fig:sum_rate_beta}.
In addition, as $P$ increases, the performance gap between the proposed DLPB scheme and the benchmark schemes grows.
This is because the proposed scheme exploits the DNN to improve the interference management at high SNRs.
Both Fig. \ref{Fig:sum_rate_power} and Fig. \ref{Fig:sum_rate_beta} show the average WSR of all users without the consideration of the CE overhead, i.e., $R = \sum_{k = 1}^K\alpha_k\mathrm{log}_2 \left(1 + \gamma_{k,t}\right)$, as defined in (\ref{SINR}).
In contrast, in Fig. \ref{Fig:sum_rate_power_protocol}, we aim to quantify the actual gain of the proposed approach by also taking into account the required CE overhead. To this end, we consider a practical system setting with carrier frequency $f_\mathrm{c} = 900~\mathrm{MHz}$ and symbol rate $T_\mathrm{s} = 66.7~\mu s$ \cite{Samsung6G}. The average velocity of the users $v$ is within $[20, 40]~\mathrm{km/h}$.
According to the transmission protocols in Fig. \ref{Fig:CE_comparison}, the time average WSRs of FP-ICSI and the proposed scheme are given by \cite{tse2005fundamentals}
$(1 - KNT_\mathrm{s}/T_\mathrm{c})R$ and $(1 - KT_\mathrm{s}/T_\mathrm{c})R$, respectively.
Here, $T_\mathrm{c} \approx \frac{1}{f_m}$ denotes the coherence time of the channel, where $f_m = \frac{v}{c}f_c$ is the Doppler frequency with $v$ and $c$ denoting the average velocity of the users and the speed of the radio waves, respectively.
Also, the neural network training set is generated based on the velocity range of $[20, 40]~\mathrm{km/h}$.
As can be observed from Fig. \ref{Fig:sum_rate_power_protocol}, if the signaling overhead is taken into account, the performance of FP-ICSI drops significantly compared with the results in Fig. \ref{Fig:sum_rate_power}.
In contrast, the proposed DLPB scheme still achieves excellent performance.
This is because the proposed scheme leverages predictive beamforming to significantly reduce the signaling overhead required for CE compared with FP-ICSI such that more time is available for data transmission, which leads to a higher system performance.
On the other hand, as the average velocity of the users increases, $T_c$ decreases and thus the performance of FP-ICSI decreases significantly, as indicated in Fig. \ref{Fig:sum_rate_power_protocol}.
In contrast, the performance of the proposed scheme remains almost the same for different velocities.
This is expected since the proposed DL-based method can adaptively learn the different features of different mobility scenarios to accurately predict the future IRS phase-shift matrix for beamforming design.
For ease of study, in the sequel, we only show the system performance without taking into account the required overhead.

%

\begin{figure}[t]
  \centering
  \includegraphics[width=3.2in,height=2.8in]{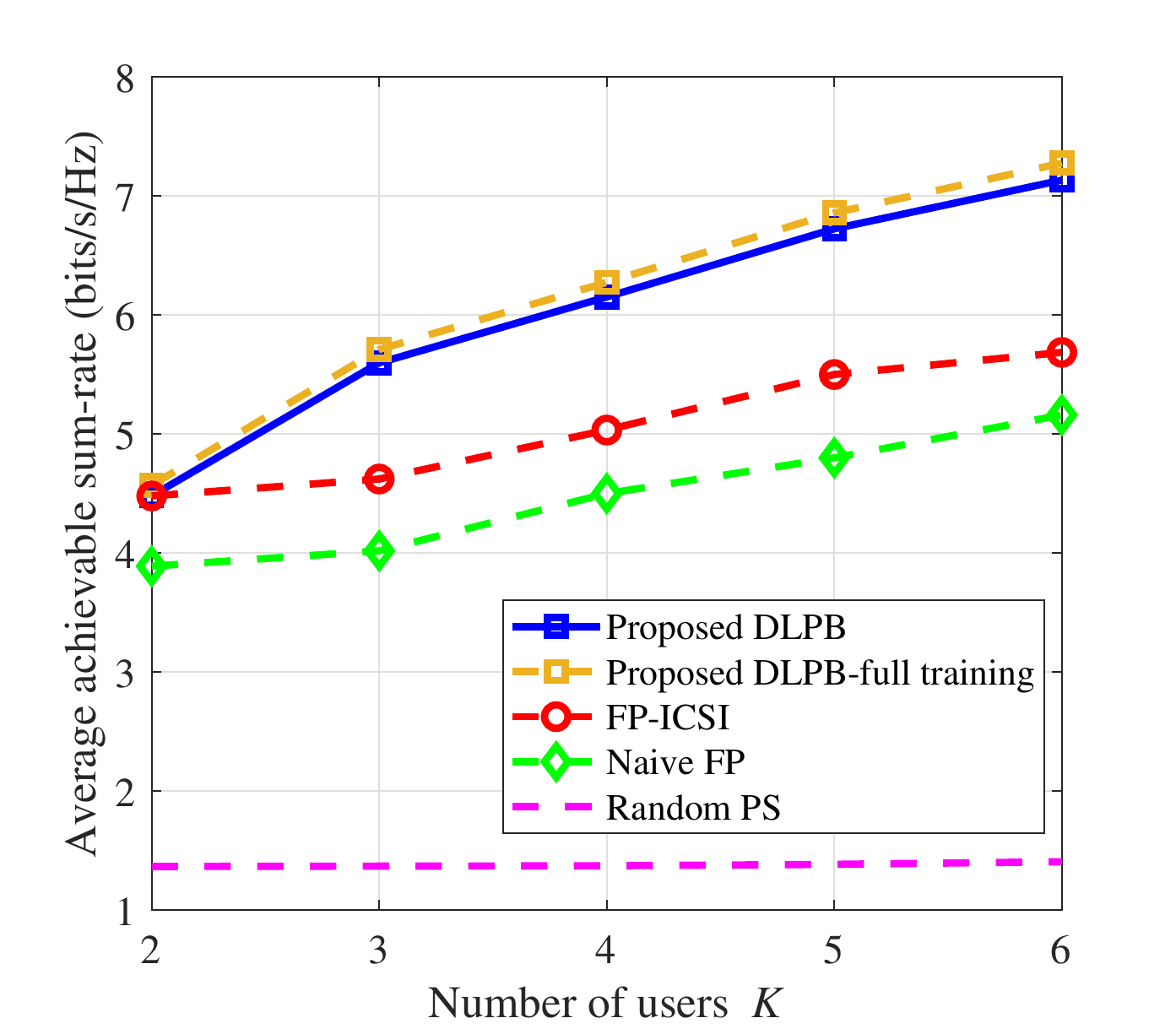}\vspace{-0.5cm}
  \caption{ Average achievable WSR versus the number of users for $\beta = 2~\mathrm{dB}$ and $P = 30~\mathrm{dBm}$.}\label{Fig:sum_rate_user}\vspace{-0.8cm}
\end{figure}

\subsection{Scalability}
To evaluate the scalability of the proposed method, in Fig. \ref{Fig:sum_rate_user}, we investigate the WSR performance for different numbers of users.
In the simulation, for the proposed DLPB method, the LA-CLGNN is trained for $K = 3$ users, and then, we directly adopt the well-trained LA-CLGNN to generate the predictive phase-shift matrix for systems with $K=\{2,3,4,5,6\}$.
To provide more insight, we also plot the results of the proposed DLPB method with full training, where the neural network is trained with the full training set generated for $K=\{2,3,4,5,6\}$.
Fig. \ref{Fig:sum_rate_user} reveals that the average WSR performance of all considered algorithms improves as $K$ increases.
Also, it can be observed that the proposed method outperforms the benchmark schemes of FP-ICSI, naive FP, and random PS, and its performance approaches that of the DLPB-full training scheme.
We note that FP-ICSI, naive FP, and random PS have to re-optimize the beamforming matrices if $K$ changes, which is complicated and introduces a large computation overhead.
In contrast, in the proposed DLPB scheme, the GNN allows each user to effectively learn how to manage the interference for an arbitrary number of users. This result shows that the proposed algorithm enjoys high scalability and is applicable for systems with different numbers of users.

\begin{figure}[t]
  \centering
  \includegraphics[width=3.2in,height=2.8in]{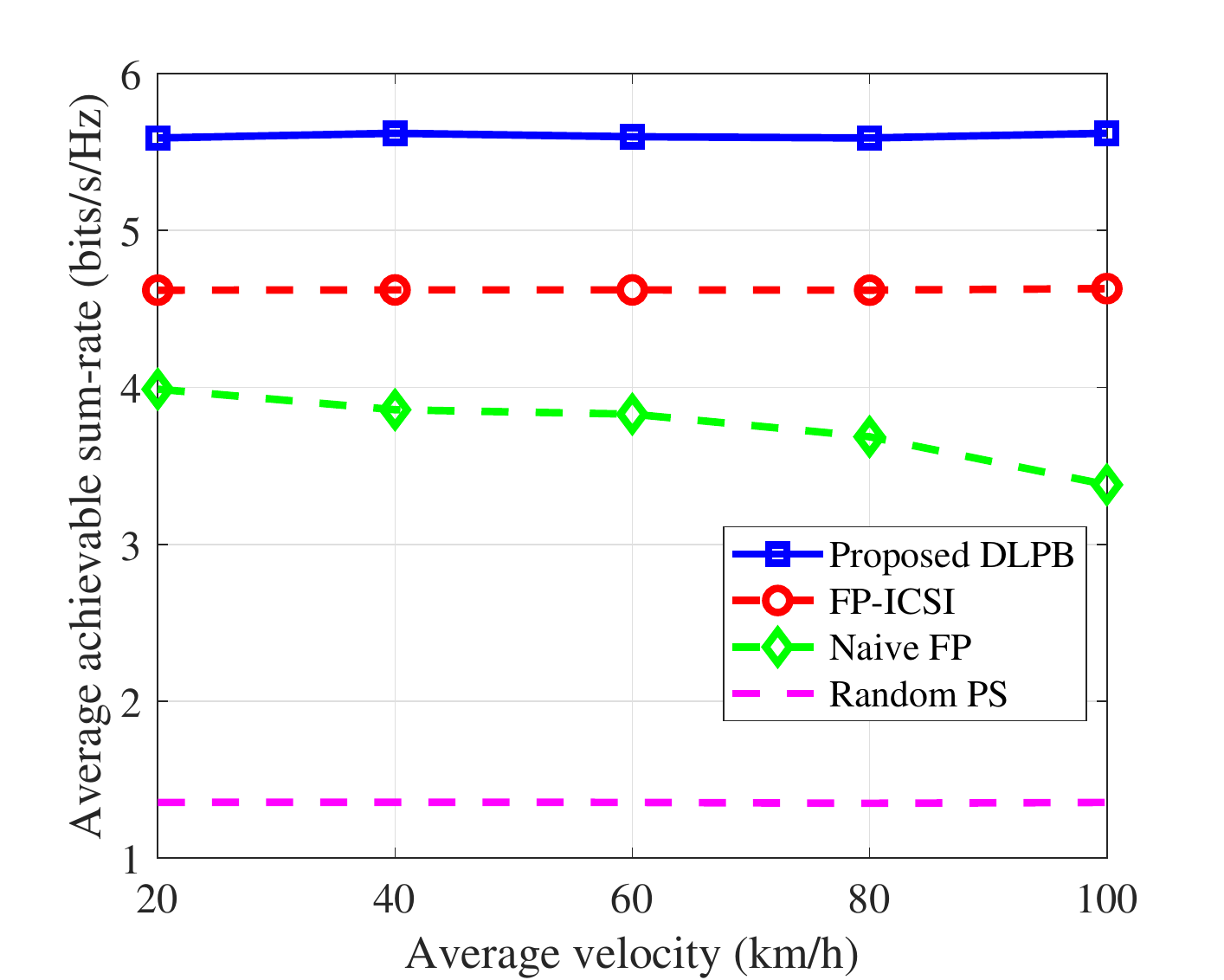}\vspace{-0.5cm}
  \caption{ Average achievable WSR versus average velocity of user for $\beta = 2~\mathrm{dB}$ and $P = 30~\mathrm{dBm}$.}\label{Fig:sum_rate_velocity}\vspace{-0.8cm}
\end{figure}

\vspace{-0.2cm}
\subsection{Generalizability}
In this section, we investigate the impact of the users' motion parameters on the WSR performance to evaluate the generalizability of the proposed method with respect to user mobility.
Fig. \ref{Fig:sum_rate_velocity} shows the average achievable WSR for different average velocities ranging from $10~\mathrm{km/h}$ to $60~\mathrm{km/h}$.
As can be seen, the WSR achieved by the naive FP method decreases as the average velocity increases.
This is due to the fact that the naive scheme directly adopts the past CSI as ICSI for beamforming design in the current time slot, and the difference between the current ICSI and the past CSI increases with the velocity.
In contrast, through the neural network training in the velocity range of $[10,60]~\mathrm{km/h}$, the proposed method leverages DL to implicitly and accurately predict features of the future LoS channels for predictive beamforming design in the next time slot.
Besides, FP-ICSI and random PS yield practically the same WSRs for different velocities, since they do not exploit any temporal information regarding the channels for beamforming design.

\begin{figure}[t]
  \centering
  \includegraphics[width=3.2in,height=2.8in]{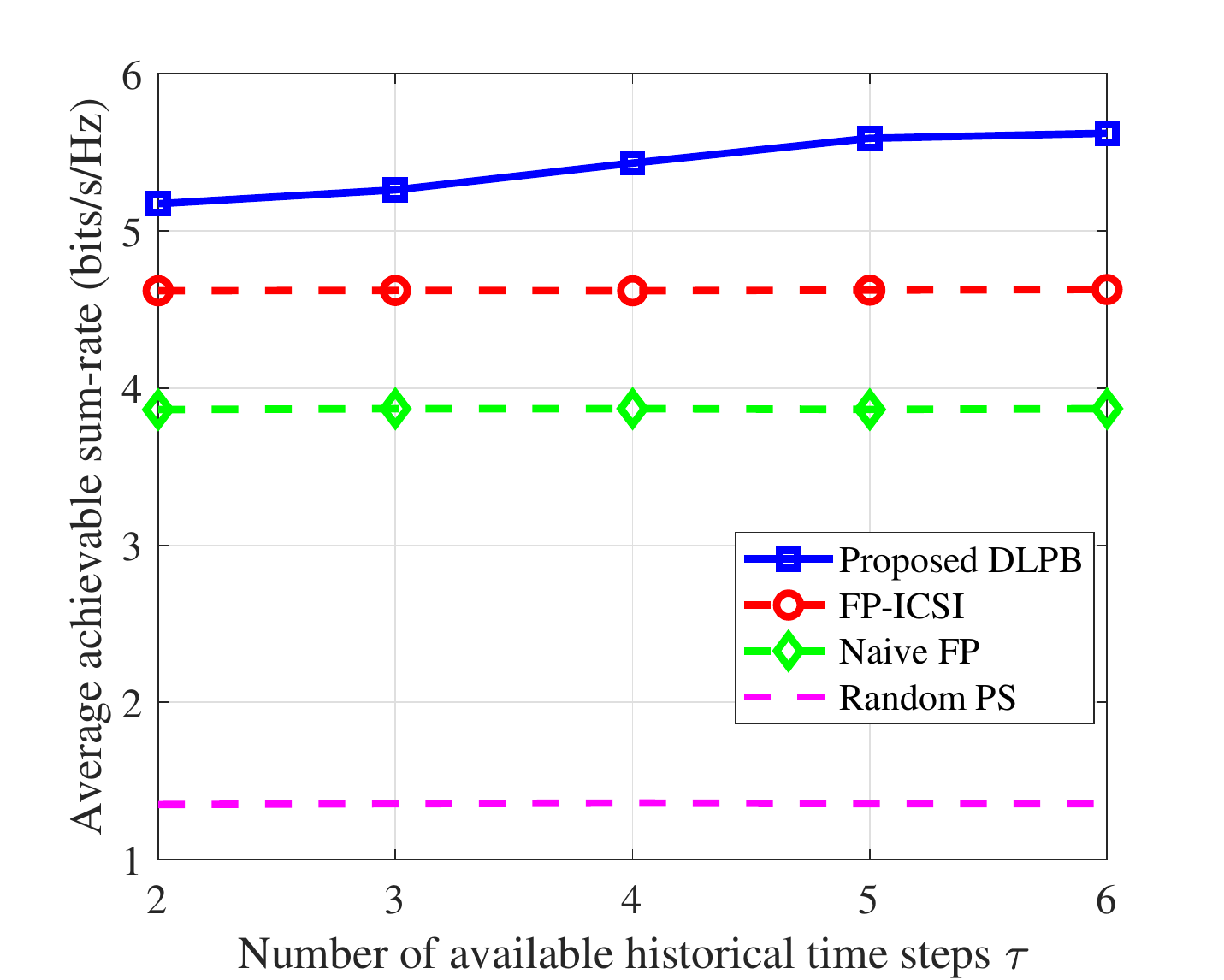}\vspace{-0.5cm}
  \caption{ Average achievable WSR versus the number of available historical time slots for $\beta = 2~\mathrm{dB}$ and $P = 30~\mathrm{dBm}$.}\label{Fig:sum_rate_tau}\vspace{-0.8cm}
\end{figure}

Next, in Fig. \ref{Fig:sum_rate_tau}, we investigate the impact of the number of available historical time steps $\tau$ on the system average WSR.
We observe that when $\tau$ increases, the WSR achieved by FP-ICSI, naive FP, and random PS remains constant.
This is expected since all these three schemes do not exploit temporal channel information for beamforming design.
However, the average WSR of the proposed DLPB scheme increases with $\tau$. This is because for large $\tau$, more historical channels can be utilized to exploit temporal dependencies for improving system performance. Yet, the additional gains become marginal for large values of $\tau$ since most useful temporal information for beamforming design is captured by the most recent LoS channel realizations.

\vspace{-0.2cm}
\section{Conclusion}
In this paper, we proposed a DL-based scalable predictive beamforming scheme for IRS-MUC systems.
In particular, a two-stage DL-based predictive beamforming framework was developed, which can directly predict the IRS beamforming matrix to bypass the need for full ICSI knowledge to reduce the CE overhead.
As a realization of the developed framework, an LA-CLGNN was designed for predictive IRS beamforming.
In LA-CLGNN, a CLSTM module was adopted to exploit the temporal dependencies from historical channels to improve the learning performance and a GNN was specifically designed to make the proposed scheme highly scalable.
Then, given the predicted IRS phase shifts, an IA-FNN was designed to optimize the instantaneous beamforming matrix at the AP.
Finally, extensive simulation results verified the excellent performance of the proposed framework in terms of WSR, scalability, and generalizability, respectively.


\bibliographystyle{ieeetr}

\setlength{\baselineskip}{10pt}

\bibliography{ReferenceSCI2}

\begin{thebibliography}{10}

\bibitem{liu2021deeplearningconference}
C.~Liu, X.~Liu, Z.~Wei, S.~Hu, D.~W.~K. Ng, and J.~Yuan, ``Deep
  learning-empowered predictive beamforming for {IRS}-assisted multi-user
  communications,'' in {\em Proc. IEEE Global Telecommun. Conf. (GLOBECOM)},
  pp.~1--6, Madrid, Spain, Dec. 2021.

\bibitem{yu2021smart}
X.~Yu, V.~Jamali, D.~Xu, D.~W.~K. Ng, and R.~Schober, ``Smart and
  reconfigurable wireless communications: From {IRS} modeling to algorithm
  design,'' {\em IEEE Wireless Commun.}, vol.~28, no.~6, pp.~118--125, Dec.
  2021.

\bibitem{liu2021reconfigurable}
Y.~Liu, X.~Liu, X.~Mu, T.~Hou, J.~Xu, M.~Di~Renzo, and N.~Al-Dhahir,
  ``Reconfigurable intelligent surfaces: Principles and opportunities,'' {\em
  IEEE Commun. Surveys Tuts.}, vol.~23, no.~3, pp.~1546--1577, 3rd Quart. 2021.

\bibitem{gong2020toward}
S.~Gong, X.~Lu, D.~T. Hoang, D.~Niyato, L.~Shu, D.~I. Kim, and Y.-C. Liang,
  ``Toward smart wireless communications via intelligent reflecting surfaces: A
  contemporary survey,'' {\em IEEE Commun. Surveys Tuts.}, vol.~22, no.~4,
  pp.~2283--2314, 4th~Quart., 2020.

\bibitem{li2021many}
D.~Li, ``How many reflecting elements are needed for energy-and
  spectral-efficient intelligent reflecting surface-assisted communication,''
  {\em IEEE Trans.\ Commun.}, vol.~70, no.~2, pp.~1320--1331, Feb. 2022.

\bibitem{wu2021intelligent}
Q.~Wu, S.~Zhang, B.~Zheng, C.~You, and R.~Zhang, ``Intelligent reflecting
  surface aided wireless communications: A tutorial,'' {\em IEEE Trans.\
  Commun.}, vol.~69, no.~5, pp.~3313--3351, May 2021.

\bibitem{di2020smart}
M.~Di~Renzo, A.~Zappone, M.~Debbah, M.-S. Alouini, C.~Yuen, J.~De~Rosny, and
  S.~Tretyakov, ``Smart radio environments empowered by reconfigurable
  intelligent surfaces: How it works, state of research, and the road ahead,''
  {\em IEEE J.\ Sel.\ Areas\ Commun.}, vol.~38, no.~11, pp.~2450--2525, Nov.
  2020.

\bibitem{pan2021reconfigurable}
C.~Pan, H.~Ren, K.~Wang, J.~F. Kolb, M.~Elkashlan, M.~Chen, M.~Di~Renzo,
  Y.~Hao, J.~Wang, A.~L. Swindlehurst, {\em et~al.}, ``Reconfigurable
  intelligent surfaces for {6G} systems: Principles, applications, and research
  directions,'' {\em IEEE Comm. Mag.}, vol.~59, no.~6, pp.~14--20, Jul. 2021.

\bibitem{yu2021convolutional}
X.~Yu, D.~Li, Y.~Xu, and Y.-C. Liang, ``Convolutional autoencoder-based phase
  shift feedback compression for intelligent reflecting surface-assisted
  wireless systems,'' {\em IEEE Commun. Lett.}, vol.~26, no.~1, pp.~89--93,
  2021.

\bibitem{Samsung6G}
Samaung, ``The next hyper-connected experience for all,'' 2020, Available:
  https://news.samsung.com/global/samsung-unveils-6g-spectrum-white-paper-and-6g-research-findings.

\bibitem{li2021intelligent}
H.~Li, W.~Cai, Y.~Liu, M.~Li, Q.~Liu, and Q.~Wu, ``Intelligent reflecting
  surface enhanced wideband {MIMO-OFDM} communications: From practical model to
  reflection optimization,'' {\em IEEE Trans.\ Commun.}, vol.~20, no.~2,
  pp.~4807--4820, Jul. 2021, [Early Access].

\bibitem{yuan2021reconfigurable}
X.~Yuan, Y.-J.~A. Zhang, Y.~Shi, W.~Yan, and H.~Liu,
  ``Reconfigurable-intelligent-surface empowered wireless communications:
  Challenges and opportunities,'' {\em IEEE Wireless Commun.}, vol.~28, no.~2,
  pp.~136--143, Apr. 2021.

\bibitem{wu2019beamforming}
Q.~Wu and R.~Zhang, ``Beamforming optimization for wireless network aided by
  intelligent reflecting surface with discrete phase shifts,'' {\em IEEE
  Trans.\ Commun.}, vol.~68, no.~3, pp.~1838--1851, Mar. 2020.

\bibitem{guo2020weighted}
H.~Guo, Y.-C. Liang, J.~Chen, and E.~G. Larsson, ``Weighted sum-rate
  maximization for reconfigurable intelligent surface aided wireless
  networks,'' {\em IEEE Trans.\ Wireless Commun.}, vol.~19, no.~5,
  pp.~3064--3076, May~2020.

\bibitem{hu2021robust}
S.~Hu, Z.~Wei, Y.~Cai, C.~Liu, D.~W.~K. Ng, and J.~Yuan, ``Robust and secure
  sum-rate maximization for multiuser {MISO} downlink systems with
  self-sustainable {IRS},'' {\em IEEE Trans. Commun.}, vol.~69, no.~10,
  pp.~7032--7049, Oct. 2021.

\bibitem{cai2022resource}
Y.~Cai, Z.~Wei, S.~Hu, C.~Liu, D.~W.~K. Ng, and J.~Yuan, ``Resource allocation
  and {3D} trajectory design for power-efficient {IRS-assisted UAV-NOMA}
  communications,'' {\em IEEE Trans. Wireless Commun.}, Jun. 2022, [Early
  Access] DOI: 10.1109/TWC.2022.3183300.

\bibitem{wang2021interplay}
J.~Wang, W.~Tang, Y.~Han, S.~Jin, X.~Li, C.-K. Wen, Q.~Cheng, and T.~J. Cui,
  ``Interplay between {RIS and AI} in wireless communications: Fundamentals,
  architectures, applications, and open research problems,'' {\em IEEE J.\
  Sel.\ Areas\ Commun.}, vol.~39, no.~8, pp.~2271--2288, Aug. 2021.

\bibitem{huang2020reconfigurable}
C.~Huang, R.~Mo, and C.~Yuen, ``Reconfigurable intelligent surface assisted
  multiuser {MISO} systems exploiting deep reinforcement learning,'' {\em IEEE
  J.\ Sel.\ Areas\ Commun.}, vol.~38, no.~8, pp.~1839--1850, Aug. 2020.

\bibitem{gao2020unsupervised}
J.~Gao, C.~Zhong, X.~Chen, H.~Lin, and Z.~Zhang, ``Unsupervised learning for
  passive beamforming,'' {\em IEEE Commun. Lett.}, vol.~24, no.~5,
  pp.~1052--1056, May 2020.

\bibitem{zheng2022survey}
B.~Zheng, C.~You, W.~Mei, and R.~Zhang, ``A survey on channel estimation and
  practical passive beamforming design for intelligent reflecting surface aided
  wireless communications,'' {\em IEEE Commun. Surveys Tuts.}, vol.~24, no.~2,
  pp.~1035--1071, 2nd Quart. 2022.

\bibitem{liu2022deepresidual}
C.~Liu, X.~Liu, D.~W.~K. Ng, and J.~Yuan, ``Deep residual learning for channel
  estimation in intelligent reflecting surface-assisted multi-user
  communications,'' {\em IEEE Trans. Wireless Commun.}, vol.~21, no.~2,
  pp.~898--912, Feb. 2022.

\bibitem{jensen2020optimal}
T.~L. Jensen and E.~De~Carvalho, ``An optimal channel estimation scheme for
  intelligent reflecting surfaces based on a minimum variance unbiased
  estimator,'' in {\em Proc. IEEE Int. Conf. Acoust. Speech Signal Process.},
  pp.~5000--5004, Barcelona,~Spain, May~2020.

\bibitem{zheng2020intelligent}
B.~Zheng and R.~Zhang, ``Intelligent reflecting surface-enhanced {OFDM}:
  Channel estimation and reflection optimization,'' {\em IEEE Wireless Commun.
  Lett.}, vol.~9, no.~4, pp.~518--522, Apr.~2020.

\bibitem{he2019cascaded}
Z.-Q. He and X.~Yuan, ``Cascaded channel estimation for large intelligent
  metasurface assisted massive {MIMO},'' {\em IEEE Wireless Commun. Lett.},
  vol.~9, no.~2, pp.~210--214, Feb.~2020.

\bibitem{tse2005fundamentals}
D.~Tse and P.~Viswanath, {\em Fundamentals of wireless communication}.
\newblock Cambridge University Press, 2005.

\bibitem{liu2020deeptransfer}
C.~Liu, Z.~Wei, D.~W.~K. Ng, J.~Yuan, and Y.-C. Liang, ``Deep transfer learning
  for signal detection in ambient backscatter communications,'' {\em IEEE
  Trans. Wireless Commun.}, vol.~20, no.~3, pp.~1624--1638, Mar. 2021.

\bibitem{lxm2020deepresidual}
X.~Liu, C.~Liu, Y.~Li, B.~Vucetic, and D.~W.~K. Ng, ``Deep residual
  learning-assisted channel estimation in ambient backscatter communications,''
  {\em IEEE Wireless Commun. Lett.}, vol.~10, no.~2, pp.~339--343, Feb.~2021.

\bibitem{yuan2020learning}
W.~Yuan, C.~Liu, F.~Liu, S.~Li, and D.~W.~K. Ng, ``Learning-based predictive
  beamforming for {UAV} communications with jittering,'' {\em IEEE Wireless
  Commun. Lett.}, vol.~9, no.~11, pp.~1970--1974, Nov.~2020.

\bibitem{liu2020location}
C.~Liu, W.~Yuan, Z.~Wei, X.~Liu, and D.~W.~K. Ng, ``Location-aware predictive
  beamforming for {UAV} communications: A deep learning approach,'' {\em IEEE
  Wireless Commun. Lett.}, vol.~10, no.~3, pp.~668--672, Mar.~2021.

\bibitem{xie2019activity}
J.~Xie, C.~Liu, Y.-C. Liang, and J.~Fang, ``Activity pattern aware spectrum
  sensing: A {CNN}-based deep learning approach,'' {\em IEEE Commun. Lett.},
  vol.~23, no.~6, pp.~1025--1028, Jun.~2019.

\bibitem{xie2020unsupervised}
J.~Xie, J.~Fang, C.~Liu, and L.~Yang, ``Unsupervised deep spectrum sensing: A
  variational auto-encoder based approach,'' {\em IEEE Trans. Veh. Technol.},
  vol.~69, no.~5, pp.~5307--5319, May 2020.

\bibitem{xie2020deep}
J.~Xie, J.~Fang, C.~Liu, and X.~Li, ``Deep learning-based spectrum sensing in
  cognitive radio: A {CNN-LSTM} approach,'' {\em IEEE Commun. Lett.}, vol.~24,
  no.~10, pp.~2196--2200, Oct.~2020.

\bibitem{wu2020comprehensive}
Z.~Wu, S.~Pan, F.~Chen, G.~Long, C.~Zhang, and S.~Y. Philip, ``A comprehensive
  survey on graph neural networks,'' {\em IEEE Trans. Neural Netw. Learn.
  Syst.}, vol.~32, no.~1, pp.~4--24, Mar. 2021.

\bibitem{shen2020graph}
Y.~Shen, Y.~Shi, J.~Zhang, and K.~B. Letaief, ``Graph neural networks for
  scalable radio resource management: Architecture design and theoretical
  analysis,'' {\em IEEE J.\ Sel.\ Areas\ Commun.}, vol.~39, no.~1,
  pp.~101--115, Jan. 2021.

\bibitem{eisen2020optimal}
M.~Eisen and A.~Ribeiro, ``Optimal wireless resource allocation with random
  edge graph neural networks,'' {\em IEEE Trans.\ Signal Proc.}, vol.~68,
  pp.~2977--2991, Apr. 2020.

\bibitem{lee2020graph}
M.~Lee, G.~Yu, and G.~Y. Li, ``Graph embedding-based wireless link scheduling
  with few training samples,'' {\em IEEE Trans.\ Wireless Commun.}, vol.~20,
  no.~4, pp.~2282--2294, Apr. 2021.

\bibitem{liu2022learning}
C.~Liu, W.~Yuan, S.~Li, X.~Liu, H.~Li, D.~W.~K. Ng, and Y.~Li, ``Learning-based
  predictive beamforming for integrated sensing and communication in vehicular
  networks,'' {\em IEEE J. Sel. Areas Commun.}, vol.~40, no.~8, pp.~2317--2334,
  Aug. 2022.

\bibitem{li2022novel}
S.~Li, W.~Yuan, C.~Liu, Z.~Wei, J.~Yuan, B.~Bai, and D.~W.~K. Ng, ``A novel
  {ISAC} transmission framework based on spatially-spread orthogonal time
  frequency space modulation,'' {\em IEEE J. Sel. Areas Commun.}, vol.~40,
  no.~6, pp.~1854--1872, Jun. 2022.

\bibitem{9076668}
W.~{Yuan}, S.~{Li}, L.~{Xiang}, and D.~W.~K. {Ng}, ``Distributed estimation
  framework for beyond {5G} intelligent vehicular networks,'' {\em IEEE Open J.
  Veh. Technol.}, vol.~1, pp.~190--214, Apr. 2020.

\bibitem{liu2014maximum}
C.~Liu and M.~Jin, ``Maximum-minimum spatial spectrum detection for cognitive
  radio using parasitic antenna arrays,'' in {\em Proc. IEEE Int. Conf. Commun.
  China (ICCC)}, pp.~365--369, Shanghai, China, Oct. 2014.

\bibitem{you2020energy}
L.~You, J.~Xiong, D.~W.~K. Ng, C.~Yuen, W.~Wang, and X.~Gao, ``Energy
  efficiency and spectral efficiency tradeoff in {RIS}-aided multiuser {MIMO}
  uplink transmission,'' {\em IEEE Trans.\ Signal Proc.}, vol.~69, pp.~1407 --
  1421, Dec. 2020.

\bibitem{liu2016blind}
C.~Liu, H.~Li, and M.~Jin, ``Blind central-symmetry-based feature detection for
  spatial spectrum sensing,'' {\em IEEE Trans.\ Veh.\ Technol.}, vol.~65,
  no.~12, pp.~10147--10152, Dec.~2016.

\bibitem{liu2015blind}
C.~Liu, M.~Li, and M.-L. Jin, ``Blind energy-based detection for spatial
  spectrum sensing,'' {\em IEEE Wireless Commun. Lett.}, vol.~4, no.~1,
  pp.~98--101, Feb.~2015.

\bibitem{li2019wirelessly}
X.~Li, G.~Zhu, Y.~Gong, and K.~Huang, ``Wirelessly powered data aggregation for
  {IoT} via over-the-air function computation: Beamforming and power control,''
  {\em IEEE Trans.\ Wireless Commun.}, vol.~18, no.~7, pp.~3437--3452, Jul.
  2019.

\bibitem{shen2012distributed}
C.~Shen, T.-H. Chang, K.-Y. Wang, Z.~Qiu, and C.-Y. Chi, ``Distributed robust
  multicell coordinated beamforming with imperfect {CSI}: An {ADMM} approach,''
  {\em IEEE Trans.\ Signal Proc.}, vol.~60, no.~6, pp.~2988--3003, Jun. 2012.

\bibitem{liu2019deep}
C.~Liu, J.~Wang, X.~Liu, and Y.-C. Liang, ``Deep {CM-CNN} for spectrum sensing
  in cognitive radio,'' {\em IEEE J. Sel. Areas Commun.}, vol.~37, no.~10,
  pp.~2306--2321, Oct.~2019.

\end{thebibliography}

\end{document}